\documentstyle[11pt]{article}
\linethickness{0.4pt}
\newcommand{\beq}{\begin{equation}}
\newcommand{\enq}{\end{equation}}

\newcommand{\beqa}{\begin{eqnarray}}
\newcommand{\enqa}{\end{eqnarray}}

\newcommand{\al}{\alpha}
\newcommand{\cH}{\cal H}
\begin{document}

\begin{titlepage}
\begin{flushright}
                Preprint IFT UWr 912/97\\
                June 1997 \\
\end{flushright}
\bigskip
\bigskip
\begin{center}
{\huge \bf Non-Critical Light-Cone String}
\end{center}
\bigskip
\bigskip
\begin{center}
{\bf  Marcin Daszkiewicz, Zbigniew Hasiewicz, Zbigniew Jask\'{o}lski}
\end{center}
\bigskip
\bigskip
\begin{center}
Institute of Theoretical Physics\\
University of Wroc{\l}aw\\
pl. Maxa Borna 9\\
PL - 50 - 206 Wroc{\l}aw
\end{center}
\vfill
\date{June 1997}
\begin{abstract}
The free non-critical string quantum model is
constructed directly in the light-cone variables
in the range of dimensions $1<D<25$.
The longitudinal degrees of freedom
are described by an abstract Verma module. The central
charge of this module is restricted by
the requirement of the closure of the nonlinear realization
of the Poincare algebra.
The spin content of the model is analysed.
In particular for $D=4$ the explicit formulae
for the character generating functions of
the open and closed massive strings are given and
the spin spectrum of first 12 excited levels is calculated.
It is shown that for the space-time dimension in
the range $1<D<25$ the non-critical light-cone string is
equivalent to the critical massive string and
to the non-critical Nambu-Goto string.
\end{abstract} 
\thispagestyle{empty}
\end{titlepage}

\section*{Introduction}

The light-cone description
of string dynamics was first analysed long time ago in
the fundamental  paper
of Goddard, Goldstone, Rebbi and Thorn \cite{ggrt},
where the covariant quantum  Nambu-Goto string model
was compared to the quantum model obtained by canonical
quantization of light-cone gauge fixed classical theory.
It was shown that the light-cone quantum
theory was relativistically invariant
  only in the critical space-time dimension
$D=26$ and for the unit intercept
of the leading Regge trajectory $\alpha(0)=1$.
On the other hand the covariant scheme of quantization
yields a ghost-free quantum theory in the space-time dimensions
$D\leq 26$ \cite{bgt,brower}.
Generically the gauge symmetry breaks down upon quantization and
in the subcritical dimensions the quantum theory
inevitably contains longitudinal degrees of freedom. In consequence
one has more degrees of freedom than in the classical model.
Only for $D=26$ and $\alpha(0)=1$
the classical symmetry is restored in the sense that
the states with longitudinal excitations become null and
drop out from physical amplitudes.
As far as the free theory is concerned
the equivalence of both method of quantization
 can be shown
for the critical dimension  and intercept
using the results of \cite{grt}.

The light-cone formalism plays a prominent role in the critical string
theory.
This is the only known formulation of  quantum string theory
in terms of physical degrees of freedom.
This is also the only framework in which one can explicitly realize
the geometric idea of joining-splitting interaction
\cite{mandelstam}. Finally
the $S$-matrix defined peturbatively within the light-cone approach
is automatically unitary. The price one has to pay for this
is a non-trivial realization of the Poincare symmetry and a difficult
and
technically demanding proof of the  Poincare invariance of
the light-cone $S$-matrix
\cite{mandelstam,sin}.
It should be stressed however that known proofs of unitarity
of the covariant theory are based either on the
equivalence of the covariant Polyakov
string theory  in the critical dimension
$D=26$ with the light-cone Mandelstam theory
\cite{wogi}, or on the so called $\alpha$-invariance in the
"covariantized light-cone" BRST formulation \cite{hikko}, which also implies
Poincare invariance of the standard light-cone formulation
\cite{kugo}.

The aim of the present paper is to show that the light-cone
formulation can be extended to non-critical string models.
As it was mentioned above such a formulation is essential
for constructing and analysing joining-splitting interaction
and a unitary perturbative expansion of the $S$-matrix.
All hitherto attempts to construct interaction
of  non-critical strings were made within
the covariant formulation of dual models \cite{chodos,mar,tho,kak}.
In the case of non-critical Polyakov string 
a straightforard extension of the dual model
conformal field theory construction
failed due to the so called  $c=1$ barrier \cite{sei}.
The equivalence of the covariant Polyakov formulation
to the Mandelstam light-cone theory is known only
for the critical dimension and intercept \cite{wogi}.
This is therefore an open question how the
$c=1$ barrier manifests itself in an
explicitly unitary formulation and
whether the joining-spliting interaction
in such a formulation
may lead to a Lorentz covariant theory
in subcritical dimensions.

In the present paper
we shall restrict ourselves to
a detailed analysis of the free theory
leaving the main problem of interactions
for future publications.
The content of the paper is as follows.
In Section 1 we construct a quantum free non-critical string
model directly in the light cone variables.
In this construction the longitudinal degrees of freedom
are described by an abstract Verma module. The central
charge of this module is restricted by
the requirement of the closure of the non-linear realization
of the Poincare algebra. Let us note that
a similar  construction was first considered by
Marnelius in the context of non-critical Polyakov string
\cite{marnelius}.
A peculiar feature of this quantum model is that
the corresponding classical system is not Lorentz covariant
because of anomalous terms in  Poisson bracket
algebra of classical Lorentz generators. These terms
are cancelled in the quantum theory by terms arising
from normal ordering and one gets a unitary
representation of the Poincare algebra. The cancellation takes
place only for  specific critical values of parameters
of this construction.

In Section 2 we analyse the spectrum of the non-critical light-cone
string introduced in Section 1. The formula for the character
generating function is obtained and its explicit expansion in
terms of irreducible characters is derived in the physical space-time
dimensions $D=4$. These results are illustrated by
numerical calculations of the spin content of the first 12 excited levels.
The corresponding results for the closed
massive strings are briefly presented in Section 3.

The rest of the paper is devoted to the relations between
various non-critical free string models. We prove
equivalence results which imply
 that the light-cone model of non-critical string
constructed in Section 1 is universal to some extend.

In Section 4 we briefly recall the massive string model.
As a quantum model it
was first constructed long time ago by Chodos and Thorn
\cite{chodos}.
It was later considered by Marnelius in the context of
non-critical Polakov string \cite{marnelius}. The relation between
these two models was also analysed in \cite{wy}.  A detailed discussion of
the classical and quantum theory  of the free massive string was 
recently given in \cite{my}.

The classical world sheet action for the massive string is
the extension of the BDHP action by the fre Liouville
action for an additional dimensionless scalar field:
\begin{eqnarray*}
S[M,g,\varphi, x] \;=& -&\!\!{\alpha\over 2\pi} \int_M \sqrt{-g}
d^2z\; g^{ab}\partial_a x^\mu \partial_b x_\mu\\
&-&\!\! {\beta\over 2\pi} \int_M \sqrt{-g}
d^2z\;\left( g^{ab}\partial_a \varphi \partial_b \varphi +
2R_g \varphi\right)\;\;\;.
\end{eqnarray*}
One of the important differences with respect to the classical
Nambu-Goto or BDHP string models is that
 the Virasoro constraints are of second class
 in the conformal gauge.
This means that one cannot use the classical light-cone
gauge in order to parameterize the reduced phase space
of the model. In fact solving classical constraints is
equivalent to solving a special class of non-linear Hill's
equations which makes canonical quantization
in the physical variables prohibitively difficult.

Using an alternative covariant method of quantization in which
constraints are imposed as conditions for physical states
one obtains a family of quantum models characterized by
the dimensionless coupling $\beta$ and the physical intercept 
$\alpha(0)$.
Among all the values of
these parameters allowed by the no-ghost theorem
the so called critical ones
$
\beta_{\rm crit} = {25-D\over 48} \;,\;
\alpha(0)_{\rm crit}  =   {D-1\over 24}
$
are especially interesting.
In this case the space $\cH$ of physical states
contains  the largest
subspace ${\cH}_0$ of null states.
In order to analyse a physical content of the model a
tractable parameterization of the quotient
\beq
\pi : {\cH} \longrightarrow {\cH}_{\rm phys}\;\; = {\cH}/{\cH}_0
\label{fibrat}
\enq
is required. One possible approach is to find a subspace
${\cH}_{\rm gauge}\subset {\cH}$ such that the projection
$\pi$ restricted to ${\cH}_{\rm gauge}$ is a 1-1 map. Such a
subspace provides a section of the fibration (\ref{fibrat})
and will be called a (quantum) gauge. Indeed the shift
by a null state can be regarded as a quantum gauge transformation
and the subspace ${\cH}_{\rm gauge}$ can be seen as a gauge slice
for this symmetry.

The basic idea in our proof of equivalence of different
string models is to regard them as different descriptions
of the critical massive string related to different quantum gauges.
In particular in Section 5
we use the quantum light-cone gauge
to prove the equivalence of the critical massive
string with the non-critical light-cone string of Section 1.
Our method is a simplification and a generalisation of the method
used in a similar context in the critical string theory
\cite{grt}.
 In Section 6 we consider another gauge slice
 which is stable with respect to the Poincare group action.
It provides a simple proof of the old conjecture \cite{marnelius}
that the critical massive string is equivalent
to the non-critical Nambu-Goto and Polyakov strings.

Finally for the sake of completeness some detailes of the DDF construction
are presented in Appendix.

\section{Non-Critical Light-Cone String}

In this section we construct a quantum
string model
in the range of space-time dimensions $1<D<25$
directly in the light-cone variables.

The starting point of the construction is a choice of a
light-cone frame in the $D$-dimensional flat Minkowski
target space.
It consists of a
pair $k,k'$ of light-like vectors $k^2=0=k'^2$
satisfying $k\cdot k' = -1$,
and a complementary set of transverse vectors
$\{e_i\}_{i=1}^{D-2}$  forming an orthonormal basis of the
subspace orthogonal to both $k$ and $k'$.
Let us denote by
\begin{eqnarray*}
{P}^+&=&k\cdot P\;\;,\;\;{x}^-=k'\cdot x \;\;,\\
P^i& =&e^i\cdot P \;\;,\;\; x^i= e^i\cdot x\;\;,\;\; i = 1,..,D-2
 \;\;,
\end{eqnarray*}
the self-adjoint
operators corresponding to the light-cone components of
momentum and position and satisfying the standard commutation relations
$$
[P^{i},x^j] \;\;  =\;\;  i\delta^{ij}\;\;\;,\;\;\;
[P^+,x^-] \;\;  =\;\;  -i\;\;\;.
$$
For each $p^+, \overline p = \sum p^i e^i$ we define
the Fock space ${\cal F}^{\rm \scriptscriptstyle T}(p^+,\overline p)$
generated by the infinite algebra
of transverse excitation operators:
$$
[\alpha_m^i,\alpha_n^j]  =  m\delta^{ij}\delta_{m,-n}\;\;\;,
\;\;\;{(\alpha_m^i)}^{\dagger}\; =\; \alpha_{-m}^i\;\;,\;\;
m,n\in { Z\!\!\!Z}
\;\;,
$$
out of the unique vacuum state $\Omega(p^+,\overline p)$ satisfying
\begin{eqnarray*}
\alpha_0^i\Omega(p^+,\overline p)& =&
{\textstyle {1\over \sqrt{\alpha}}}p^i\Omega(p^+,\overline p) \;\;\;,\\
\alpha_0^+\Omega(p^+,\overline p)& =&
{\textstyle {1\over \sqrt{\alpha}}}p^+\Omega(p^+,\overline p) \;\;\;,\\
\alpha_m^i\Omega(p^+,\overline p)& = & 0
\;\;\;,\;\;\;m > 0\;\;\;.
\end{eqnarray*}
We use the standard relation $\al_0^\mu ={1\over \sqrt{\alpha}} P^\mu$,
with the  dimensionful parameter $\alpha$ related to the conventional
Regge slope $\alpha'$  by $\alpha = {1\over 2 \alpha'}$.
For later convenience we introduce
the transverse Virasoro generators
\beq
L_n^{\rm \scriptscriptstyle T}
=  {\textstyle\frac{1}{2}}\sum_{m=-\infty}^{+\infty}
: \overline{\alpha}_{-m}\cdot \overline{\alpha}_{n+m} :
\;\;\;,\label{ltra}
\enq
satisfying the algebra:
$$
[L_n^{\rm \scriptscriptstyle T},L_m^{\rm \scriptscriptstyle T}] =
(n-m)L_{n+m}^{\rm \scriptscriptstyle T} + {\textstyle{D-2\over 12}}  (n^3 - n)
\delta_{n,-m}\;\;\;.
$$
In the non-critical
light-cone string model longitudinal excitations are described by
the Verma module ${\cal V}^{\rm \scriptscriptstyle L}(b)$ generated
by the Virasoro algebra
\begin{equation}
{[L^{\rm \scriptscriptstyle L}_n,L^{\rm \scriptscriptstyle L}_m]}=
(n-m)L^{\rm \scriptscriptstyle L}_{n+m} +
{\textstyle {c\over 12}} (n^3-n)\delta_{m,-n}
\label{verma}
\end{equation}
out of the highest wight state
$$
L^{\rm \scriptscriptstyle L}_0\: \Omega^{\rm \scriptscriptstyle L}(b)
 = b\:\Omega(b)\;\;\;,\;\;\;
L^{\rm \scriptscriptstyle L}_{n}\: \Omega^{\rm \scriptscriptstyle L}(b)
\;=\; 0\;\;,\;\;n>0
\;\;\;.
$$
For the central charge $c$ of this algebra in the range $1<c<25$ and
for $b>0$
the hermicity properties of the generators:
$$
{L^{\rm \scriptscriptstyle L}_n}^\dagger
= L^{\rm \scriptscriptstyle L}_{-n}
\;\;,\;\;n\in Z\!\!\!Z\;\;\;,
$$
determine a positively defined non-degenerate inner
product inducing a Hilbert space structure on
${\cal V}^{\rm \scriptscriptstyle L}(b)$. For $b=0$
this inner product acquires null directions
and for $b<0$ one gets negative norm states in
${\cal V}^{\rm \scriptscriptstyle L}(b)$.

The full space of states in the noncritical light-cone string model
is defined as the direct integral of Hilbert spaces
$$
{H}_{\rm lc} = \int_{I\!\!R \setminus \{0\}} {dp_+ \over | p^+|}
 \int_{I\!\!R^{d-2}} d^{d-2}\overline{p}\;
 {\cal F}^{\rm \scriptscriptstyle T}(p^+,\overline p)\otimes
 {\cal V}^{\rm \scriptscriptstyle L}(b)\;\;\;.
$$
In order to complete the construction one has
to introduce a unitary realization
of the Poincare algebra on ${H}_{\rm lc}$.
The generators of translations in the longitudinal and the
transverse directions are given by
the operators $P^+$ and $P^i, i=1,..., d-1$, respectively.
The generator of translation in the $x^+$-direction  is
defined by the self-adjoint operator
\beq
{P}^- =
{\frac{\alpha}{P^+}} (L_0^{\rm \scriptscriptstyle T} +
L_0^{\rm \scriptscriptstyle L}- a_0) \;\;\;.
\label{lctr}
\enq
Within the light-cone formulation the
$x^+$ coordinate is regarded as an evolution parameter.
In consequence $P^-$ plays the role
of the Hamiltonian and the Schr\"{o}dinger equation
reads
$$
i {\partial \over \partial x^+} \Psi =
P^- \Psi \;\;\;.
$$
The generators of Lorentz rotations are defined
by the self-adjoint operators
\beqa
{M}^{ij}_{\rm lc\;\;} &=& {P}^i{x}^j-{P}^j{x}^i +
i\sum_{n\geq 1}
{\frac{1}{n}} (\al_{-n}^i\al_{n}^j - \al_{-n}^j\al_{n}^i)
\;\;\;,
\nonumber \\
{M}^{i+}_{\rm lc\;\;} &=& {P}^+ {x}^i \;\;\;,\nonumber\\
{M}^{+-}_{\rm lc\;\;} &=& {\textstyle\frac{1}{2}}({P}^+{x}^-+x^-P^+) \;\;\;,
\label{rot} \\
{M}^{i-}_{\rm lc\;\;} &=& {\textstyle\frac{1}{2}}( {x}^i P^-
 + P^- {x}^i ) - {P}^i{x}^-\nonumber\\
& -&\;i{\frac{ \sqrt{\alpha}}{P^+}} \sum_{n\geq 1}
{\frac{1}{n}}
\left(
\al_{-n}^i(L_n^{\rm \scriptscriptstyle T}
+L_n^{\rm \scriptscriptstyle L} ) -
(L_{-n}^{\rm \scriptscriptstyle T}
+L_{-n}^{\rm \scriptscriptstyle L}) \al_n^i \right),
\nonumber
\enqa
The algebra of the generators $P^+,P^i$, $P^-$ (\ref{lctr}), and
$M^{\mu\nu}$ (\ref{rot}) closes to the Lie algebra of Poincare
group if and only if the central charge $c$ of the Virasoro
algebra generating the "longitudinal" Verma module
${\cal V}^{\rm \scriptscriptstyle L}(b)$
and $a_0$ entering the definition of the Hamiltonian
of the system  (\ref{lctr}) take the critical values
$$
c = 26-D\;\;\;,\;\;\; a_0 = 1\;\;\;.
$$
Indeed, tedious but straightforward calculations show that
the only anomalous terms appear in the commutators:
\beqa
\left[\; {M}^{i-}_{\rm lc}\;,\;{M}^{j-}_{\rm lc}\right]& =&
-\left ( 2- {\textstyle {D-2+c \over 12}} \right )
\frac{\alpha}{{P^+}^2}\sum_{n>0} n
(\al_{-n}^i\al_{n}^j - \al_{-n}^j\al_{n}^i) \nonumber\\
& & -\left( {\textstyle {D-2 +c \over 12}} - 2a_0 \right )
\frac{\alpha}{{P^+}^2}\sum_{n>0} {1\over n}
(\al_{-n}^i\al_{n}^j - \al_{-n}^j\al_{n}^i)\;\;\;. \nonumber
\enqa
Let us note that the model still contains one free parameter $b$
entering the mass shell condition
\begin{equation}
M^2 =2P^+P^- - \overline{P}^2\;=\;2\alpha\left(
R_{\rm lc} +b - 1\right)\;\;\;,
\label{mass}
\end{equation}
where
\begin{equation}
R_{\rm lc} =
L_0^{\rm \scriptscriptstyle T} -
{\textstyle{1\over 2}}\overline{\alpha}^2_0
+ L_0^{\rm \scriptscriptstyle L} - b
\label{level}
\end{equation}
is the light-cone level operator.
The only restriction on $b$ is
$b\geq 0$. As it was mentioned above this
is a necessary and sufficient condition for
the absence of ghost
states in the "longitudinal" Verma module with the central
charge $c = 26-D$. 

In order to analyse the
spin content of the model it is convenient
to work with the explicit Fock space realization \cite{chodos,nove}
of the "longitudinal" Verma module which can be constructed
for $b\geq {25-D\over 24}$.
Let  us denote by ${\cal F}^{\rm \scriptscriptstyle L}(q)$
the Fock space  generated by the infinite algebra
of "Liouville" excitation operators:
$$
[\beta_m,\beta_n] =  m\delta_{m,-n} \;\;\;
,\;\;\;(\beta_m)^\dagger \;=\; \beta_{-m}\;\;\;,\;\;\;
m,n\in { Z\!\!\!Z} \;\;\;,
$$
out of the unique vacuum state $\Omega(q)$ satisfying
\begin{eqnarray*}
\beta_0\Omega(q)& =& q\Omega(q) \;\;,\;\;
\beta_m\Omega(q)\; =\; 0\;\;,\;m > 0\;\;\;.
\end{eqnarray*}
For  each  $n\in Z\!\!\!Z $ and a dimensionless real parameter $\beta$
we define
\beq
L_{n}^{\rm \scriptscriptstyle L} =
 {\textstyle\frac{1}{2}}\sum_{k= -\infty}^{+\infty}
 :\beta_{-k}\beta_{n+k}: +
2{\sqrt\beta}in\beta_{n} + 2\beta\delta_{n,0}\;\;\;,
\label{lob}
\enq
satisfying
\begin{eqnarray*}
[L_n^{\rm \scriptscriptstyle L},L_m^{\rm \scriptscriptstyle L}]
&=& (n-m)L_{n+m}^{\rm \scriptscriptstyle L} +
{\textstyle {1+48\beta\over 12}} (n^3 - n)
\delta_{n,-m}\;\;\;,\\
{L_n^{\rm \scriptscriptstyle L}}^\dagger &=& L_{-n}^{\rm \scriptscriptstyle L}\;\;\;,\\
L_0^{\rm \scriptscriptstyle L}\Omega(q) &=&
({\textstyle{1\over 2}} {q}^2 + 2\beta)
\Omega(q)\;\;\;.
\end{eqnarray*}
It follows that for $\beta = {25-D\over 48}$, and
${q}= \sqrt{2b-{25-D\over 12}}$ the
operators (\ref{lob}) can be identified with the Virasoro generators
of the "longitudinal" Verma module (\ref{verma}) and the spaces
${\cal F}^{\rm \scriptscriptstyle L}(q)$,
and ${\cal V}^{\rm \scriptscriptstyle L}(b)$ are isomorphic with respect to
the inner product structure.

The generators of the Poincare algebra
regarded as operators on the Fock space
\begin{equation}
{H}_{\rm lc} = \int_{I\!\!R \setminus \{0\}} {dp_+ \over | p^+|}
 \int_{I\!\!R^{d-2}} d^{d-2}\overline{p}\;
  {\cal F}^{\rm \scriptscriptstyle T}(p^+,\overline p)\otimes
 {\cal F}^{\rm \scriptscriptstyle L}(q)
 \;\;\;,        \label{ssss}
\end{equation}
are uniquely determined by their action on
the vacuum states $\Omega(p^+,\overline p)\otimes
\Omega({q})$ and the commutation relations
with the excitation operators:
\beqa
\left [{P}^-,\alpha_n^i \right ] & = & -\alpha \frac {n}{p^+}
\alpha_n^i \;\;\;,
\;\;\;\;
\left [{P}^-,\beta_n \right ] \;=\; -\alpha\frac
{n}{p^+} \beta_n \;\;\; ,\cr
\left [{M}^{ij}_{\rm lc}\;,\alpha_n^k \right ] & = & -i(\alpha_n^j
\delta^{ik} - \alpha_m^i \delta^{jk})~~\;\;\;, \cr
\left [{M}^{i-}_{\rm lc},\beta_n \right ] & = &
-\alpha\frac {n x^i}{p^+} \beta_n
+ \sqrt\alpha \frac {in}{p^+} \left (\sum_{m>0} \frac {1}{m}
(\alpha_{-m}^i \beta_{n+m} - \alpha_m^i\beta_{n-m}) \right )  \cr
&  &-{\sqrt\alpha}\frac {2}{p^+} n \sqrt {\beta} \alpha _n^i \;\;\;,
\label{lclocom}\\
\left [{M}^{i-}_{\rm lc},\alpha_n^j \right ] & = & -\alpha\frac {nx^i}{p^+}
\alpha_n^j + {\sqrt\alpha} \frac {in}{p^+} \left (\sum_{m>0}
\frac{1}{m}
(\alpha_{-m}^i\alpha_{m+n}^j - \alpha_{n-m}^j\alpha_m^i) \right )   \cr
&  & +{\sqrt{\alpha}} \frac {i}{p^+} \delta^{ij}
\left( L_n^{\rm \scriptscriptstyle T}+L_n^{\rm \scriptscriptstyle L}
\right)\;\;\;,
\nonumber
\enqa
with all remaining commutators being zero.

\section{Little Group SO$(D-1)$ and Spin Content}

In this section we shall analyse the spin content of the noncritical
light-cone string
using the Fock space realization. Our results
are therefore limited to the models with the parameter
$b$ in the range $b\geq {25-D\over 24}$.
Since the generators of the Poincare algebra
(\ref{lctr},\ref{rot})
commute with the light-cone
level operator $R_{\rm lc}$ (\ref{level})
entering the mass shell condition (\ref{mass})
the decomposition of
$H_{\rm lc}$ into representations of a fixed
mass coincides with the level decomposition
$$
H_{\rm lc} =  \bigoplus_{N \geq 0} H_{\rm lc}^{(N)}\;\;\;,
\;\;\;
R_{\rm lc}\;H_{\rm lc}^{(N)}\; =\; NH_{\rm lc}^{(N)}\;\;\;.
$$
For $b$ in the range ${25-D\over24} \leq b<1$
the lowest level subspace $H_{\rm lc}^{(0)}$ carries
the irreducible tachyonic representation with $m^2=b-1$.
For $b=1$ $H_{\rm lc}^{(0)}$ is a massless, and for
$b>1$ a massive scalar representation. This completes
analysis of the spin content of the lowest level.

For the whole range ${25-D\over 24}<b$
all higher level representations
$H_{\rm lc}^{(N)}$ have real non-zero mass and can
be further decomposed into particle $H_{\rm lc}^{+(N)}$ and
antiparticle $H_{\rm lc}^{-(N)}$ representations.
We shall  find the decomposition of
$H_{\rm lc}^{+(N)}$ into irreducible
representations.

Let us fix some level $N > 0$.
For any momentum $p$
satisfying $-p^2=m^2= 2\alpha(N+b-1)$ and $p^+>0$
one can always  choose a light-cone frame $\{k,k',e^i\}$
such that
$$
p = {\sqrt \alpha}k' +
{\sqrt \alpha}\left(N + b-1 \right) k\;\;\;.
$$
In this frame the operators $ G^j ,\, j = 1...D-2  $ :
\beq
{G }^j = i{\textstyle {m^2 \over 2\alpha}} M^{j+} -
i M^{j-} =-{\sum_{n>0}}
{ {1 \over n}}
\left(\alpha_{-n}^j
 (L_{n}^{\rm \scriptscriptstyle T}+ L_{n}^{\rm \scriptscriptstyle L})
 - (L_{-n}^{\rm \scriptscriptstyle T} + L_{-n}^{\rm \scriptscriptstyle L})
 \alpha_n^j\right)\;\;\;,
\label{gene}
\enq
together with the infinitesimal transverse rotations of ${\rm SO}(D-2)$
\beq
iM^{ji} = -{\sum_{n>0}}
{1 \over n}(\alpha_{-n}^j \alpha_{n}^i - \alpha_{-n}^i \alpha_n^j)
\;\;\;,
\label{sod}
\enq
generate unitary action of the little group ${\rm SO}(D-1)$ of the momentum $p$.

One should notice that split of ${so}(D-1)$ onto (\ref{gene}) and (\ref{sod})
is of symmetric type i.e. (\ref{gene}) generate the whole Lie  algebra by
commutators and much of the analysis can be restricted to
their representation only.

Let us consider the subspace
$H_p \subset H^{+(N)}_{\rm lc}$ of
all states of level $N$ with the fixed momentum $p$ ($p^+>0$).
We shall calculate the character
\beq
\chi_N(g) = {\rm tr}{\cal D} (g)
\label{char}
\enq
of the corresponding unitary representation ${\cal D}$ of the
little group ${\rm SO}(D-1)$ on $H_p$. For this purpose
it is convenient to decompose $H_p$
according to the partitions of $N$.
We denote by ${\cal P}(N)$ the set of all
partitions of $N$.
Any partition  $p(N) \in {\cal P}(N)$ is uniquely characterized by the
sequence $p(N) = ({m_N},{m_{N-1}},...,{m_1})$ with $\sum  km_k = N$.
For every  partition $p(N)$ of $N$ we define the corresponding Hilbert
subspace   $H^{p(N)}_p\subset H_p $ spanned by all states of the form
$$
A_{-N}^{a_N^{m_N}}A_{-N+1}^{a_{N-1}^1}...
A_{-N+1}^{a_{N-1}^{m_{N-1}}}...
A_{-1}^{a_1^1}...
A_{-1}^{a_1^{m_1}}\Omega (p^+,\overline{p})\otimes
\Omega(q)
$$
where, for $m>0$
$$
A_{-m}^a = \cases{ i\beta_{-m}
\;& for $a=0\;$,\cr
\alpha _{-m}^a\;& for  $a=1,...,d-2\;$,\cr}
$$
and $q = \sqrt{2b-{25-D\over 12}}$.
\footnote{The factor of $i$ in front of $\beta_{-m}$ is inserted
in order to obtain
the standard real representation of the ${\rm so}(D-1)$ Lie algebra
in (\ref{comm}).}
Let $V_{-m}$  be the vector space spanned by the
the excitation operators $\left\{A_{-m}^a\right\}_{a=0}^{d-2}$
of level $m$. Then
\beq
H_p^{p(N)} \simeq  (\otimes^{m_N}_SV_{-N})\otimes ...
              ...\otimes (\otimes^{m_1}_SV_{-1})\;\;\;,
\label{tensor}
\enq
where $\otimes^m_S$ denotes the $m$-th symmetric tensor power.
The Hilbert space $H_p$ decomposes into orthogonal sum
\beq
H_p = \bigoplus_{p(N) \in {\cal P}(N)} H_p^{p(N)}
\label{roz}
\enq
and one has the corresponding partition of identity
\beq
id = \sum_{p(N) \in {\cal P}(N)} \pi_{p(N)}
\label{jed}
\enq
with $\{ \pi_{p(N)} \}$ being the set of projectors corresponding to (\ref{roz}).
Inserting (\ref{jed}) on both sides of ${\cal D}(g)$ in (\ref{char}) and
using ${\rm tr}(\pi_{p(N)}{\cal D}(g)\pi_{p'(N)}) = 0$ for
$p(N)\neq p'(N)$ one gets immediately that
\beq
\chi_N(g) = \sum_{p(N)\in {\cal P}(N)} {\rm tr}( \pi_{p(N)}{\cal D}(g)\pi_{p(N)})
\;\;\;.
\label{charsplit}
\enq
The virtue of the formula above is that the traces on the r.h.s.
are characters of the representations of ${\rm SO}(D-1)$ on the subspaces
(\ref{tensor}), regarded as appropriate tensor products
of the fundamental vector representations on the $(D-1)$-dimensional
vector spaces $V_{-m}, m>0$.
This is evident on the Lie algebra level. Infinitesimal action
of ${\cal D}(g)$ on $H_p$  (\ref{tensor})
with $g$ generated by (\ref{gene})
is given by the sum of commutators of (\ref{gene}) with
the fundamental excitation operators:
\beq
\left[\; G^i \;,\; A^a_{-m} \right]\;=\;
    2m {\sqrt \beta} D_{\; \;b}^{(i) \; a}A^b_{-m} +
    M^{(i) a}(m)\;\;\;,
\label{comm}
\enq
where $D^{(i)}$ are the standard antisymmetric matrices of fundamental
representation of ${ so}(D-1)$ with $1$ on $i$-th place in the first row and
zero elsewhere. The exact form of the operators $M^{(i) a}(m)$ can be
easily obtained from (\ref{lclocom}). One should only take
into account that
in the kinematical situation assumed at the
beginning of this section
they are of second
order in the excitation operators.
Their contributions to the infinitesimal
actions of ${\rm SO}(D-1)$ on  $H_p^{p(N)}$
correspond therefore to different partitions of $N$ and are zero under
projection $\pi_{p(N)}$.
Consequently the operators $\pi_{p(N)}{\cal D}(g)\pi_{p(N)}$
define
corresponding tensor representations of ${\rm SO}(D-1)$
on the spaces $H_p^{p(N)}$.
On the infinitesimal level they are described
by the first term of the r.h.s. of (\ref{comm}).

Taking into account the above property of
$\pi_{p(N)}{\cal D}(g)\pi_{p(N)}$ one can rewrite
the formula (\ref{charsplit}) for  $\chi_N$ in the following form
\beq
\chi_N = \sum_{p(N)\in {\cal P}(N)}
 \prod_{m_k \in p(N)}\chi_{S}^{m_k}\;\;\;,
\label{charexp}
\enq
with $\chi_{S}^{m_k}$ being the character of $m_k$-th symmetric tensor
power of the fundamental (vector) representation of ${\rm SO}(D-1)$.

In order to describe the spin content of the model it is convenient to
introduce the "generating" function for characters:
\beq
\chi(t,g) := \sum_{N\ge 0}t^N \chi_N (g)\;\;\;.
\label{genera}
\enq
Inserting (\ref{charexp}) into (\ref{genera}) and taking into account the
formula for symmetric characters \cite{weyl} one obtains:
\beq
\chi(t,g) \; = \; \prod_{k\ge 1}{ 1\over {\rm det}(1-t^k{\cal D}_f(g))}
\;\;\;,
\label{gen}
\enq
where ${\cal D}_f$ denotes fundamental representation of ${\rm SO}(D-1)$. As the
characters are class functions the domain of (\ref{gen}) can be restricted
to the maximal torus of ${\rm SO}(D-1)$ and in fact to the fundamental domain
describing the quotient $T_{{\rm SO}(D-1)}/{\cal W}$ of the torus
by the Weyl group.

The method of finding character generating functions for critical bosonic
and fermionic strings was presented in details
in the paper \cite{curttho}.
The formulae obtained in that article
after multiplication by the partition function
$p(t)={{\prod}_{n=1}^{\infty}} {(1-t^n)}^{-1}$ give the results for massive
string.
This extra factor in front of generating functions corresponds to the fact
that compared to the critical models there is one additional family of modes
to generate massive string states.

We give the formulae for the character generating function in most
interesting case of $D=4$. This case is also technically simplest as the
maximal torus of the little group ${\rm SO}(3)$ is 1-dimensional and
the representations are labelled by one non-negative integer.
The function
(\ref{gen}) can be written as:
\beq
\chi (t,\varphi )= {p(t)}^3\sum_{k>0}\sum_{j\ge 0}
t^{\frac{k(k-1)}{2}}{(-1)}^{k-1}{(1-t^k)}^2t^{kj}\chi^j(\varphi)\;\;\;,
\label{so3ge}
\enq
where
\beq
\chi^j(\varphi)= \frac{\sin \left ((j+\frac{1}{2})\varphi \right )}
{\sin\left (\frac{\varphi}{2}\right )}\;\;\;,
\label{so3cha}
\enq
is the irreducible character of spin $j\in I\!\!N$.
The spin spectrum of  the noncritical light-cone  string
with the physical intercept
$$
\alpha(0) = b-1\; =\; {\textstyle {1\over 2}}q^2
- {\textstyle {D-1\over 24}}\; =\; - {\textstyle {D-1\over 24}}
$$
up to
the 12th excited level is presented on Fig.1. The first 4 levels
of the corresponding closed string model are also displayed for
comparison.

\begin{picture}(130.00,160.00)(-65.00,15.00)
\put(-55.875,27.00){\vector(0,1){140.00}}
\multiput(-55.875,49.67)(0.00,10.00){12}{\line(-1,0){1.00}}
\put(-60.00,49.67){\makebox(0,0)[r]{\small 1}}
\put(-60.00,59.67){\makebox(0,0)[r]{\small 2}}
\put(-60.00,69.67){\makebox(0,0)[r]{\small 3}}
\put(-60.00,79.67){\makebox(0,0)[r]{\small 4}}
\put(-60.00,89.67){\makebox(0,0)[r]{\small 5}}
\put(-60.00,99.67){\makebox(0,0)[r]{\small 6}}
\put(-60.00,109.67){\makebox(0,0)[r]{\small 7}}
\put(-60.00,119.67){\makebox(0,0)[r]{\small 8}}
\put(-60.00,129.67){\makebox(0,0)[r]{\small 9}}
\put(-60.00,139.67){\makebox(0,0)[r]{\small 10}}
\put(-60.00,149.67){\makebox(0,0)[r]{\small 11}}
\put(-60.00,159.67){\makebox(0,0)[r]{\small 12}}
\put(-70.00,39.67){\vector(1,0){140.00}}
\put(60,30.67){\makebox(0,0)[cc]{\small $m^2/2\alpha$}}
\put(-55.00,165.00){\makebox(0,0)[l]{\small spin}}
\put(3.00,25.00){\makebox(0,0)[cc]{\rm Fig.1}}
\put(-25.00,135.00){\framebox(33,22)[cc]{\shortstack{$D=4$\\
                                                 $\alpha(0)_{\rm open}
                                                 = - {1\over 8}$\\
                                                 $\alpha(0)_{\rm closed}
                                                 = - {1\over 4}$
          }}}
\put(-20.00,127.00){\circle*{2.00}}
\put(-18.00,127.00){\makebox(0,0)[l]{\small open string}}
\put(-20.00,122.00){\circle{2.00}}
\put(-18.00,122.00){\makebox(0,0)[l]{\small closed string}}
\put(-57.00,39.67){\circle*{2.00}}
\put(-47.00,49.67){\circle*{2.00}}
\put(-37.00,39.67){\circle*{2.00}}
\put(-37.00,49.67){\circle*{2.00}}
\put(-37.00,59.67){\circle*{2.00}}
\put(-27.00,39.67){\circle*{2.00}}
\put(-27.00,49.67){\circle*{2.00}}
\put(-27.00,59.67){\circle*{2.00}}
\put(-27.00,69.67){\circle*{2.00}}
\put(-17.00,39.67){\circle*{2.00}}
\put(-17.00,49.67){\circle*{2.00}}
\put(-17.00,59.67){\circle*{2.00}}
\put(-17.00,69.67){\circle*{2.00}}
\put(-17.00,79.67){\circle*{2.00}}
\put(-7.00,39.67){\circle*{2.00}}
\put(-7.00,49.67){\circle*{2.00}}
\put(-7.00,59.67){\circle*{2.00}}
\put(-7.00,69.67){\circle*{2.00}}
\put(-7.00,79.67){\circle*{2.00}}
\put(-7.00,89.67){\circle*{2.00}}
\put(-37,43.67){\makebox(0,0)[cc]{\small 1}}
\put(-47,53.67){\makebox(0,0)[cc]{\small 1}}
\put(-37,53.67){\makebox(0,0)[cc]{\small 1}}
\put(-37,63.67){\makebox(0,0)[cc]{\small 1}}
\put(-27,53.67){\makebox(0,0)[cc]{\small 3}}
\put(-27,73.67){\makebox(0,0)[cc]{\small 1}}
\put(-27,63.67){\makebox(0,0)[cc]{\small 1}}
\put(-27,43.67){\makebox(0,0)[cc]{\small 1}}
\put(-17,83.67){\makebox(0,0)[cc]{\small 1}}
\put(-17,73.67){\makebox(0,0)[cc]{\small 1}}
\put(-15,63.67){\makebox(0,0)[cc]{\small 4}}
\put(-15,53.67){\makebox(0,0)[cc]{\small 4}}
\put(-15,43.67){\makebox(0,0)[cc]{\small 3}}
\put(-7,93.67){\makebox(0,0)[cc]{\small 1}}
\put(-7,83.67){\makebox(0,0)[cc]{\small 1}}
\put(-7,73.67){\makebox(0,0)[cc]{\small 4}}
\put(-7,63.67){\makebox(0,0)[cc]{\small 6}}
\put(-7,53.67){\makebox(0,0)[cc]{\small 9}}
\put(-7,43.67){\makebox(0,0)[cc]{\small 3}}
\put(3,39.67){\circle*{2.00}}
\put(3,49.67){\circle*{2.00}}
\put(3,59.67){\circle*{2.00}}
\put(3,69.67){\circle*{2.00}}
\put(3,79.67){\circle*{2.00}}
\put(3,89.67){\circle*{2.00}}
\put(3,99.67){\circle*{2.00}}
\put(3,43.67){\makebox(0,0)[cc]{\small 8}}
\put(3,53.67){\makebox(0,0)[cc]{\small 13}}
\put(3,63.67){\makebox(0,0)[cc]{\small 13}}
\put(3,73.34){\makebox(0,0)[cc]{\small 7}}
\put(3,83.67){\makebox(0,0)[cc]{\small 4}}
\put(3,93.67){\makebox(0,0)[cc]{\small 1}}
\put(3,103.67){\makebox(0,0)[cc]{\small 1}}
\put(13,49.67){\circle*{2.00}}
\put(13,59.67){\circle*{2.00}}
\put(13,69.67){\circle*{2.00}}
\put(13,79.67){\circle*{2.00}}
\put(13,89.67){\circle*{2.00}}
\put(13,99.67){\circle*{2.00}}
\put(13,109.67){\circle*{2.00}}
\put(13,53.67){\makebox(0,0)[cc]{\small 25}}
\put(13,63.67){\makebox(0,0)[cc]{\small 21}}
\put(13,73.34){\makebox(0,0)[cc]{\small 15}}
\put(13,83.67){\makebox(0,0)[cc]{\small 7}}
\put(13,93.67){\makebox(0,0)[cc]{\small 4}}
\put(13,103.67){\makebox(0,0)[cc]{\small 1}}
\put(13,113.67){\makebox(0,0)[cc]{\small 1}}
\put(23,49.67){\circle*{2.00}}
\put(23,59.67){\circle*{2.00}}
\put(23,69.67){\circle*{2.00}}
\put(23,79.67){\circle*{2.00}}
\put(23,89.67){\circle*{2.00}}
\put(23,99.67){\circle*{2.00}}
\put(23,109.67){\circle*{2.00}}
\put(23,119.67){\circle*{2.00}}
\put(25,43.67){\makebox(0,0)[cc]{\small 19}}
\put(25,53.67){\makebox(0,0)[cc]{\small 37}}
\put(25,63.67){\makebox(0,0)[cc]{\small 40}}
\put(25,73.34){\makebox(0,0)[cc]{\small 25}}
\put(25,83.67){\makebox(0,0)[cc]{\small 16}}
\put(23,93.67){\makebox(0,0)[cc]{\small 7}}
\put(23,103.67){\makebox(0,0)[cc]{\small 4}}
\put(23,113.67){\makebox(0,0)[cc]{\small 1}}
\put(23,123.67){\makebox(0,0)[cc]{\small 1}}
\put(33,49.67){\circle*{2.00}}
\put(33,59.67){\circle*{2.00}}
\put(33,69.67){\circle*{2.00}}
\put(33,79.67){\circle*{2.00}}
\put(33,89.67){\circle*{2.00}}
\put(33,99.67){\circle*{2.00}}
\put(33,109.67){\circle*{2.00}}
\put(33,119.67){\circle*{2.00}}
\put(33,129.67){\circle*{2.00}}
\put(33,53.67){\makebox(0,0)[cc]{\small 65}}
\put(33,63.67){\makebox(0,0)[cc]{\small 62}}
\put(33,73.67){\makebox(0,0)[cc]{\small 49}}
\put(33,83.67){\makebox(0,0)[cc]{\small 27}}
\put(33,93.67){\makebox(0,0)[cc]{\small 16}}
\put(33,103.67){\makebox(0,0)[cc]{\small 7}}
\put(33,113.67){\makebox(0,0)[cc]{\small 4}}
\put(33,123.67){\makebox(0,0)[cc]{\small 1}}
\put(33,133.67){\makebox(0,0)[cc]{\small 1}}
\put(43,49.67){\circle*{2.00}}
\put(43,59.67){\circle*{2.00}}
\put(43,69.67){\circle*{2.00}}
\put(43,79.67){\circle*{2.00}}
\put(43,89.67){\circle*{2.00}}
\put(43,99.67){\circle*{2.00}}
\put(43,109.67){\circle*{2.00}}
\put(43,119.67){\circle*{2.00}}
\put(43,129.67){\circle*{2.00}}
\put(43,139.67){\circle*{2.00}}
\put(43,53.67){\makebox(0,0)[cc]{\small 97}}
\put(43,63.67){\makebox(0,0)[cc]{\small 109}}
\put(43,73.34){\makebox(0,0)[cc]{\small 79}}
\put(43,83.67){\makebox(0,0)[cc]{\small 53}}
\put(43,93.67){\makebox(0,0)[cc]{\small 28}}
\put(43,103.67){\makebox(0,0)[cc]{\small 16}}
\put(43,113.67){\makebox(0,0)[cc]{\small 7}}
\put(43,123.67){\makebox(0,0)[cc]{\small 4}}
\put(43,133.67){\makebox(0,0)[cc]{\small 1}}
\put(43,143.67){\makebox(0,0)[cc]{\small 1}}
\put(53,49.67){\circle*{2.00}}
\put(53,59.67){\circle*{2.00}}
\put(53,69.67){\circle*{2.00}}
\put(53,79.67){\circle*{2.00}}
\put(53,89.67){\circle*{2.00}}
\put(53,99.67){\circle*{2.00}}
\put(53,109.67){\circle*{2.00}}
\put(53,119.67){\circle*{2.00}}
\put(53,129.67){\circle*{2.00}}
\put(53,139.67){\circle*{2.00}}
\put(53,149.67){\circle*{2.00}}
\put(53,53.67){\makebox(0,0)[cc]{\small 160}}
\put(53,63.67){\makebox(0,0)[cc]{\small 169}}
\put(53,73.34){\makebox(0,0)[cc]{\small 139}}
\put(53,83.67){\makebox(0,0)[cc]{\small 88}}
\put(53,93.67){\makebox(0,0)[cc]{\small 55}}
\put(53,103.67){\makebox(0,0)[cc]{\small 28}}
\put(53,113.67){\makebox(0,0)[cc]{\small 16}}
\put(53,123.67){\makebox(0,0)[cc]{\small 7}}
\put(53,133.67){\makebox(0,0)[cc]{\small 4}}
\put(53,143.67){\makebox(0,0)[cc]{\small 1}}
\put(53,153.67){\makebox(0,0)[cc]{\small 1}}
\put(63,49.67){\circle*{2.00}}
\put(63,59.67){\circle*{2.00}}
\put(63,69.67){\circle*{2.00}}
\put(63,79.67){\circle*{2.00}}
\put(63,89.67){\circle*{2.00}}
\put(63,99.67){\circle*{2.00}}
\put(63,109.67){\circle*{2.00}}
\put(63,119.67){\circle*{2.00}}
\put(63,129.67){\circle*{2.00}}
\put(63,139.67){\circle*{2.00}}
\put(63,149.67){\circle*{2.00}}
\put(63,159.67){\circle*{2.00}}
\put(65,43.67){\makebox(0,0)[cc]{\small 105}}
\put(65,53.67){\makebox(0,0)[cc]{\small 238}}
\put(65,63.67){\makebox(0,0)[cc]{\small 277}}
\put(65,73.67){\makebox(0,0)[cc]{\small 222}}
\put(65,83.67){\makebox(0,0)[cc]{\small 157}}
\put(65,93.67){\makebox(0,0)[cc]{\small 92}}
\put(65,103.67){\makebox(0,0)[cc]{\small 56}}
\put(63,113.67){\makebox(0,0)[cc]{\small 28}}
\put(63,123.67){\makebox(0,0)[cc]{\small 16}}
\put(63,133.67){\makebox(0,0)[cc]{\small 7}}
\put(63,143.67){\makebox(0,0)[cc]{\small 4}}
\put(63,153.67){\makebox(0,0)[cc]{\small 1}}
\put(63,163.67){\makebox(0,0)[cc]{\small 1}}
\put(13,39.67){\circle*{2.00}}
\put(23,39.67){\circle*{2.00}}
\put(33,39.67){\circle*{2.00}}
\put(43,39.67){\circle*{2.00}}
\put(53,39.67){\circle*{2.00}}
\put(63,39.67){\circle*{2.00}}
\put(13,43.67){\makebox(0,0)[cc]{\small 9}}
\put(33,43.67){\makebox(0,0)[cc]{\small 25}}
\put(43,43.67){\makebox(0,0)[cc]{\small 45}}
\put(53,43.67){\makebox(0,0)[cc]{\small 61}}
\put(-58.125,39.67){\circle{2.00}}
\put(-18.125,39.67){\circle{2.00}}
\put(-18.125,49.67){\circle{2.00}}
\put(-18.125,59.67){\circle{2.00}}
\put(-20.125,36.33){\makebox(0,0)[cc]{\small 1}}
\put(-20.125,46.33){\makebox(0,0)[cc]{\small 1}}
\put(-20.125,56.33){\makebox(0,0)[cc]{\small 1}}
\put(21.825,39.67){\circle{2.00}}
\put(21.825,49.67){\circle{2.00}}
\put(21.825,59.67){\circle{2.00}}
\put(21.825,69.67){\circle{2.00}}
\put(21.825,79.67){\circle{2.00}}
\put(19.825,36.33){\makebox(0,0)[cc]{\small 3}}
\put(19.825,46.33){\makebox(0,0)[cc]{\small 6}}
\put(19.825,56.33){\makebox(0,0)[cc]{\small 6}}
\put(19.825,66.33){\makebox(0,0)[cc]{\small 3}}
\put(19.825,76.33){\makebox(0,0)[cc]{\small 1}}
\put(61.825,39.67){\circle{2.00}}
\put(61.825,49.67){\circle{2.00}}
\put(61.825,59.67){\circle{2.00}}
\put(61.825,69.67){\circle{2.00}}
\put(61.825,79.67){\circle{2.00}}
\put(61.825,89.67){\circle{2.00}}
\put(61.825,99.67){\circle{2.00}}
\put(59.825,36.33){\makebox(0,0)[cc]{\small 12}}
\put(59.825,46.33){\makebox(0,0)[cc]{\small 25}}
\put(59.825,56.33){\makebox(0,0)[cc]{\small 27}}
\put(59.825,66.33){\makebox(0,0)[cc]{\small 18}}
\put(59.825,76.33){\makebox(0,0)[cc]{\small 10}}
\put(59.825,86.33){\makebox(0,0)[cc]{\small 3}}
\put(59.825,96.33){\makebox(0,0)[cc]{\small 1}}
\end{picture}

\section{Closed Noncritical Light-Cone  String}

Follwoin standard lines the construction of Section 2 can be extended
to the closed string case following standard lines.
The closed string Hilbert space is the "diagonal"
part of the tensor product
 $$
 H_{\rm cl} = \widetilde{H}_{\rm lc}
 \otimes_{\rm \scriptscriptstyle D} {H}_{\rm lc} \;
 =\;  \bigoplus_{N \geq 0} \widetilde{H}_{\rm lc}^{(N)}
 \otimes H_{\rm lc}^{(N)}\;\;\;,
 $$
of two copies of the open string Hilbert spaces.
The canonical variables
in the sector $H_{\rm lc}$ are usually called right-movers while those
in the sector $\widetilde{H}_{\rm lc}$ - left movers.
The right and the left-movers
commute with each other by construction.
Both sectors are related by the conditions:
\begin{eqnarray*}
{\tilde \alpha}_0^{\mu}&=&
\alpha_0^{\mu}\;=\;{P^{\mu}\over 2\sqrt{\alpha}}\;\;\;, \;\;\;
\mu =0,...,D-1\;\;\;,\\
\widetilde{c}& =& c\;\;\;,\;\;\;\widetilde{b}\;=\; b\;\;\;.
\end{eqnarray*}
The Hamiltonian generating the $x^+$-evolution of the system
is given by
\beq
{P}^-_{\rm cl}=
{\frac{\alpha}{P^+}}
(\widetilde{L}_0^{\rm \scriptscriptstyle T} + L_0^{\rm \scriptscriptstyle T} +
\widetilde{L}_0^{\rm \scriptscriptstyle L} + L_0^{\rm \scriptscriptstyle L}
- 2a_0) \;\;\;.
\label{clha}
\enq
Similarly the other Poincare generators are defined by the formulae
\beqa
{M}^{ij}_{\rm lc\;\;} &=& {P}^i{x}^j-{P}^j{x}^i \nonumber \\
&&+i\sum_{n\geq 1}
{\frac{1}{n}}
(\widetilde{\al}_{-n}^i\widetilde{\al}_{n}^j -
\widetilde{\al}_{-n}^j\widetilde{\al}_{n}^i
+ \al_{-n}^i\al_{n}^j - \al_{-n}^j\al_{n}^i)
\;\;\;,
\nonumber \\
{M}^{i+}_{\rm lc\;\;} &=& {P}^+ {x}^i \;\;\;,\nonumber\\
{M}^{+-}_{\rm lc\;\;} &=& {\textstyle\frac{1}{2}}({P}^+{x}^-+x^-P^+) \;\;\;,
\label{clrot} \\
{M}^{i-}_{\rm lc\;\;} &=& {\textstyle\frac{1}{2}}( {x}^i P^-_{\rm cl}
 + P^-_{\rm cl} {x}^i ) - {P}^i{x}^-\nonumber\\
& &-i{\frac{ \sqrt{\alpha}}{P^+}} \sum_{n\geq 1}
{\frac{1}{n}}
\left(
\widetilde{\al}_{-n}^i(\widetilde{L}_n^{\rm \scriptscriptstyle T}
+\widetilde{L}_n^{\rm \scriptscriptstyle L} ) -
(\widetilde{L}_{-n}^{\rm \scriptscriptstyle T}
+\widetilde{L}_{-n}^{\rm \scriptscriptstyle L}) \widetilde{\al}_n^i
\right.\nonumber\\
&&\;\;\;\;\;\;\;\;\;\;\;\;\;\;\;\;\;\;\;\;\;\;\;\;\;\;\;
+\left.\al_{-n}^i(L_n^{\rm \scriptscriptstyle T}
+L_n^{\rm \scriptscriptstyle L} ) -
(L_{-n}^{\rm \scriptscriptstyle T}
+L_{-n}^{\rm \scriptscriptstyle L}) \al_n^i
\right)\;\;\;,
\nonumber
\enqa
The conditions for the closure of the Poincare algebra are
$a_0=1$ and the central charges in the right and the left
"longitudinal" Verma module satisfying $\widetilde{c} = c = 26-D$.

As in the case of the open string the longitudinal Verma module can be
realized in a Fock space for $b>{25-D\over 24}$.
The left and the right "Liouville" sectors are related by the
condition $\widetilde{q}=q$.
The mass shell condition reads
$$
M^2 =2P^+P^- - \overline{P}^2\;=\;4\alpha\left(
2N + \alpha(0) \right)\;\;\;,
$$
with the "physical" intercept given by
$
\alpha(0) = 2b -2 =
{q}^2 - {\textstyle{D-1\over 12}}.
$

The generating function for characters can be
identified with the "diagonal" part (i.e. all terms of the form $x^Ny^N$)
of the product of two open string generating functions.
In the case $D=4$ one has
\begin{eqnarray*}
\chi(x,y,\varphi)&=& \chi(x,\varphi)\chi(y,\varphi)\\
&=&p^3(x)p^3(y)
\sum_{k=1}^{\infty}\sum_{i=0}^{\infty}
\sum_{l=1}^{\infty}\sum_{j=0}^{\infty}
\sum_{m=|i-j|}^{i+j}\\
&&
(-1)^{k+l}x^{\frac{k(k-1)}{2}+ki} y^{\frac{l(l-1)}{2}+ lj}
(1-x^k)^2    (1-y^l)^2
\frac{\sin\left((m+\frac{1}{2})\varphi\right)}{\sin\left(\frac{\varphi}{2}
\right)}           \;\;\;.
\end{eqnarray*}
The spectrum of the closed light-cone string in 4 dimensions and
with the "physical"
intercept $\alpha(0)= -{1\over 4}$ is presented
for first 11 excited levels and up to the spin 11 on Fig.2.

\begin{picture}(190.00,145.00)(-60.00,20.00)
\put(-56.875,27.34){\vector(0,1){130.00}}
\multiput(-56.875,49.67)(0.00,10.00){11}{\line(-1,0){1.00}}
\put(-60.00,49.67){\makebox(0,0)[r]{\small 1}}
\put(-60.00,59.67){\makebox(0,0)[r]{\small 2}}
\put(-60.00,69.67){\makebox(0,0)[r]{\small 3}}
\put(-60.00,79.67){\makebox(0,0)[r]{\small 4}}
\put(-60.00,89.67){\makebox(0,0)[r]{\small 5}}
\put(-60.00,99.67){\makebox(0,0)[r]{\small 6}}
\put(-60.00,109.67){\makebox(0,0)[r]{\small 7}}
\put(-60.00,119.67){\makebox(0,0)[r]{\small 8}}
\put(-60.00,129.67){\makebox(0,0)[r]{\small 9}}
\put(-60.00,139.67){\makebox(0,0)[r]{\small 10}}
\put(-60.00,149.67){\makebox(0,0)[r]{\small 11}}
\put(60.00,30.67){\makebox(0,0)[cc]{\small $m^2/8\alpha$}}
\put(-53.00,155.00){\makebox(0,0)[l]{\small spin}}
\put(3.00,25.00){\makebox(0,0)[cc]{\rm Fig.2}}
\put(-50.00,130.00){\framebox(33,17)[cc]{\shortstack{$D=4$\\
                                                 $\alpha(0)_{\rm closed}
                                                 = - {1\over 4}$
          }}}
\put(-70.00,39.67){\vector(1,0){140.00}}
\put(-58,39.67){\circle*{2.00}}
\put(-48,39.67){\circle*{2.00}}
\put(-48,49.67){\circle*{2.00}}
\put(-48,59.67){\circle*{2.00}}
\put(-38,39.67){\circle*{2.00}}
\put(-38,49.67){\circle*{2.00}}
\put(-38,59.67){\circle*{2.00}}
\put(-38,69.67){\circle*{2.00}}
\put(-38,79.67){\circle*{2.00}}
\put(-28,39.67){\circle*{2.00}}
\put(-28,49.67){\circle*{2.00}}
\put(-28,59.67){\circle*{2.00}}
\put(-28,69.67){\circle*{2.00}}
\put(-28,79.67){\circle*{2.00}}
\put(-28,89.67){\circle*{2.00}}
\put(-28,99.67){\circle*{2.00}}
\put(-18,39.67){\circle*{2.00}}
\put(-18,49.67){\circle*{2.00}}
\put(-18,59.67){\circle*{2.00}}
\put(-18,69.67){\circle*{2.00}}
\put(-18,79.67){\circle*{2.00}}
\put(-18,89.67){\circle*{2.00}}
\put(-18,99.67){\circle*{2.00}}
\put(-18,109.67){\circle*{2.00}}
\put(-18,119.67){\circle*{2.00}}
\put(-8,39.67){\circle*{2.00}}
\put(-8,49.67){\circle*{2.00}}
\put(-8,59.67){\circle*{2.00}}
\put(-8,69.67){\circle*{2.00}}
\put(-8,79.67){\circle*{2.00}}
\put(-8,89.67){\circle*{2.00}}
\put(-8,99.67){\circle*{2.00}}
\put(-8,109.67){\circle*{2.00}}
\put(-8,119.67){\circle*{2.00}}
\put(-8,129.67){\circle*{2.00}}
\put(-8,139.67){\circle*{2.00}}
\put(-48,43.67){\makebox(0,0)[cc]{\small 1}}
\put(-48,53.67){\makebox(0,0)[cc]{\small 1}}
\put(-48,63.67){\makebox(0,0)[cc]{\small 1}}
\put(-38,43.67){\makebox(0,0)[cc]{\small 3}}
\put(-38,83.67){\makebox(0,0)[cc]{\small 1}}
\put(-38,53.67){\makebox(0,0)[cc]{\small 6}}
\put(-38,63.67){\makebox(0,0)[cc]{\small 6}}
\put(-38,73.67){\makebox(0,0)[cc]{\small 3}}
\put(-28,103.67){\makebox(0,0)[cc]{\small 1}}
\put(-28,93.67){\makebox(0,0)[cc]{\small 3}}
\put(-28,53.67){\makebox(0,0)[cc]{\small 25}}
\put(-28,83.67){\makebox(0,0)[cc]{\small 10}}
\put(-28,73.67){\makebox(0,0)[cc]{\small 18}}
\put(-28,63.67){\makebox(0,0)[cc]{\small 27}}
\put(-28,43.67){\makebox(0,0)[cc]{\small 12}}
\put(-18,123.67){\makebox(0,0)[cc]{\small 1}}
\put(-18,113.67){\makebox(0,0)[cc]{\small 3}}
\put(-18,103.67){\makebox(0,0)[cc]{\small 12}}
\put(-18,93.67){\makebox(0,0)[cc]{\small 28}}
\put(-18,83.67){\makebox(0,0)[cc]{\small 58}}
\put(-18,73.67){\makebox(0,0)[cc]{\small 90}}
\put(-18,63.67){\makebox(0,0)[cc]{\small 116}}
\put(-18,53.67){\makebox(0,0)[cc]{\small 100}}
\put(-18,43.67){\makebox(0,0)[cc]{\small 43}}
\put(-8,143.67){\makebox(0,0)[cc]{\small 1}}
\put(-8,133.67){\makebox(0,0)[cc]{\small 3}}
\put(-8,123.67){\makebox(0,0)[cc]{\small 12}}
\put(-8,113.67){\makebox(0,0)[cc]{\small 32}}
\put(-8,103.67){\makebox(0,0)[cc]{\small 78}}
\put(-8,93.67){\makebox(0,0)[cc]{\small 150}}
\put(-8,83.67){\makebox(0,0)[cc]{\small 258}}
\put(-8,73.34){\makebox(0,0)[cc]{\small 366}}
\put(-8,63.67){\makebox(0,0)[cc]{\small 429}}
\put(-8,53.67){\makebox(0,0)[cc]{\small 355}}
\put(-8,43.67){\makebox(0,0)[cc]{\small 144}}
\put(2,39.67){\circle*{2.00}}
\put(2,49.67){\circle*{2.00}}
\put(2,59.67){\circle*{2.00}}
\put(2,69.67){\circle*{2.00}}
\put(2,79.67){\circle*{2.00}}
\put(2,89.67){\circle*{2.00}}
\put(2,99.67){\circle*{2.00}}
\put(2,109.67){\circle*{2.00}}
\put(2,119.67){\circle*{2.00}}
\put(2,129.67){\circle*{2.00}}
\put(2,139.67){\circle*{2.00}}
\put(2,149.67){\circle*{2.00}}
\put(2,43.67){\makebox(0,0)[cc]{\small 469}}
\put(2,53.67){\makebox(0,0)[cc]{\small 1199}}
\put(2,63.67){\makebox(0,0)[cc]{\small 1507}}
\put(2,73.34){\makebox(0,0)[cc]{\small 1386}}
\put(2,83.67){\makebox(0,0)[cc]{\small 1052}}
\put(2,93.67){\makebox(0,0)[cc]{\small 679}}
\put(2,103.67){\makebox(0,0)[cc]{\small 393}}
\put(2,113.67){\makebox(0,0)[cc]{\small 198}}
\put(2,123.67){\makebox(0,0)[cc]{\small 90}}
\put(2,133.67){\makebox(0,0)[cc]{\small 34}}
\put(2,143.67){\makebox(0,0)[cc]{\small 12}}
\put(2,153.67){\makebox(0,0)[cc]{\small 3}}
\put(12,39.67){\circle*{2.00}}
\put(12,49.67){\circle*{2.00}}
\put(12,59.67){\circle*{2.00}}
\put(12,69.67){\circle*{2.00}}
\put(12,79.67){\circle*{2.00}}
\put(12,89.67){\circle*{2.00}}
\put(12,99.67){\circle*{2.00}}
\put(12,109.67){\circle*{2.00}}
\put(12,119.67){\circle*{2.00}}
\put(12,129.67){\circle*{2.00}}
\put(12,139.67){\circle*{2.00}}
\put(12,149.67){\circle*{2.00}}
\put(12,43.67){\makebox(0,0)[cc]{\small 1439}}
\put(12,53.67){\makebox(0,0)[cc]{\small 3764}}
\put(12,63.67){\makebox(0,0)[cc]{\small 4878}}
\put(12,73.67){\makebox(0,0)[cc]{\small 4707}}
\put(12,83.67){\makebox(0,0)[cc]{\small 3785}}
\put(12,93.67){\makebox(0,0)[cc]{\small 2632}}
\put(12,103.67){\makebox(0,0)[cc]{\small 1648}}
\put(12,113.67){\makebox(0,0)[cc]{\small 929}}
\put(12,123.67){\makebox(0,0)[cc]{\small 483}}
\put(12,133.67){\makebox(0,0)[cc]{\small 222}}
\put(12,143.67){\makebox(0,0)[cc]{\small 94}}
\put(12,153.67){\makebox(0,0)[cc]{\small 34}}
\put(22,39.67){\circle*{2.00}}
\put(22,49.67){\circle*{2.00}}
\put(22,59.67){\circle*{2.00}}
\put(22,69.67){\circle*{2.00}}
\put(22,79.67){\circle*{2.00}}
\put(22,89.67){\circle*{2.00}}
\put(22,99.67){\circle*{2.00}}
\put(22,109.67){\circle*{2.00}}
\put(22,119.67){\circle*{2.00}}
\put(22,129.67){\circle*{2.00}}
\put(22,139.67){\circle*{2.00}}
\put(22,149.67){\circle*{2.00}}
\put(22,43.67){\makebox(0,0)[cc]{\small 4278}}
\put(22,53.67){\makebox(0,0)[cc]{\small 11373}}
\put(22,63.67){\makebox(0,0)[cc]{\small 15117}}
\put(22,73.67){\makebox(0,0)[cc]{\small 15168}}
\put(22,83.67){\makebox(0,0)[cc]{\small 12786}}
\put(22,93.67){\makebox(0,0)[cc]{\small 9420}}
\put(22,103.67){\makebox(0,0)[cc]{\small 6276}}
\put(22,113.67){\makebox(0,0)[cc]{\small 3813}}
\put(22,123.67){\makebox(0,0)[cc]{\small 2157}}
\put(22,133.67){\makebox(0,0)[cc]{\small 1119}}
\put(22,143.67){\makebox(0,0)[cc]{\small 541}}
\put(22,153.67){\makebox(0,0)[cc]{\small 234}}
\put(32,39.67){\circle*{2.00}}
\put(32,49.67){\circle*{2.00}}
\put(32,59.67){\circle*{2.00}}
\put(32,69.67){\circle*{2.00}}
\put(32,79.67){\circle*{2.00}}
\put(32,89.67){\circle*{2.00}}
\put(32,99.67){\circle*{2.00}}
\put(32,109.67){\circle*{2.00}}
\put(32,119.67){\circle*{2.00}}
\put(32,129.67){\circle*{2.00}}
\put(32,139.67){\circle*{2.00}}
\put(32,149.67){\circle*{2.00}}
\put(32,43.67){\makebox(0,0)[cc]{\small 12147}}
\put(32,53.67){\makebox(0,0)[cc]{\small 32708}}
\put(32,63.67){\makebox(0,0)[cc]{\small 44372}}
\put(32,73.67){\makebox(0,0)[cc]{\small 45939}}
\put(32,83.67){\makebox(0,0)[cc]{\small 40205}}
\put(32,93.67){\makebox(0,0)[cc]{\small 30999}}
\put(32,103.67){\makebox(0,0)[cc]{\small 21705}}
\put(32,113.67){\makebox(0,0)[cc]{\small 13980}}
\put(32,123.67){\makebox(0,0)[cc]{\small 8420}}
\put(32,133.67){\makebox(0,0)[cc]{\small 4735}}
\put(32,143.67){\makebox(0,0)[cc]{\small 2509}}
\put(32,153.67){\makebox(0,0)[cc]{\small 1229}}
\put(42,39.67){\circle*{2.00}}
\put(42,49.67){\circle*{2.00}}
\put(42,59.67){\circle*{2.00}}
\put(42,69.67){\circle*{2.00}}
\put(42,79.67){\circle*{2.00}}
\put(42,89.67){\circle*{2.00}}
\put(42,99.67){\circle*{2.00}}
\put(42,109.67){\circle*{2.00}}
\put(42,119.67){\circle*{2.00}}
\put(42,129.67){\circle*{2.00}}
\put(42,139.67){\circle*{2.00}}
\put(42,149.67){\circle*{2.00}}
\put(43,43.67){\makebox(0,0)[cc]{\small 33472}}
\put(43,53.67){\makebox(0,0)[cc]{\small 91073}}
\put(43,63.67){\makebox(0,0)[cc]{\small 125695}}
\put(43,73.67){\makebox(0,0)[cc]{\small 133512}}
\put(43,83.67){\makebox(0,0)[cc]{\small 120564}}
\put(43,93.67){\makebox(0,0)[cc]{\small 96531}}
\put(43,103.67){\makebox(0,0)[cc]{\small 70441}}
\put(43,113.67){\makebox(0,0)[cc]{\small 47556}}
\put(43,123.67){\makebox(0,0)[cc]{\small 30116}}
\put(43,133.67){\makebox(0,0)[cc]{\small 17961}}
\put(43,143.67){\makebox(0,0)[cc]{\small 10163}}
\put(43,153.67){\makebox(0,0)[cc]{\small 5421}}
\put(52,39.67){\circle*{2.00}}
\put(52,49.67){\circle*{2.00}}
\put(52,59.67){\circle*{2.00}}
\put(52,69.67){\circle*{2.00}}
\put(52,79.67){\circle*{2.00}}
\put(52,89.67){\circle*{2.00}}
\put(52,99.67){\circle*{2.00}}
\put(52,109.67){\circle*{2.00}}
\put(52,119.67){\circle*{2.00}}
\put(52,129.67){\circle*{2.00}}
\put(52,139.67){\circle*{2.00}}
\put(52,149.67){\circle*{2.00}}
\put(54,43.67){\makebox(0,0)[cc]{\small 89079}}
\put(54,53.67){\makebox(0,0)[cc]{\small 244350}}
\put(54,63.67){\makebox(0,0)[cc]{\small 342192}}
\put(54,73.67){\makebox(0,0)[cc]{\small 371322}}
\put(54,83.67){\makebox(0,0)[cc]{\small 344292}}
\put(54,93.67){\makebox(0,0)[cc]{\small 284409}}
\put(54,103.67){\makebox(0,0)[cc]{\small 214857}}
\put(54,113.67){\makebox(0,0)[cc]{\small 150774}}
\put(54,123.67){\makebox(0,0)[cc]{\small 99558}}
\put(54,133.67){\makebox(0,0)[cc]{\small 62232}}
\put(54,143.67){\makebox(0,0)[cc]{\small 37074}}
\put(54,153.67){\makebox(0,0)[cc]{\small 21027}}
\end{picture}

\section{Critical massive string}

The quantum massive string model is 
formulated in terms of the pseudo-Hilbert space defined \cite{my} as a direct
integral of Fock spaces $H_p$:
\beq
 H = \int d^Dp~H_p\;\;\;.
\label{ss}
\enq
The integration
ranges over D-dimensional spectrum of the self-adjoint momentum operators
$\{ {P}^\mu ; \mu = 0,..,D-1 \}$, which together with their canonical
conjugates satisfy standard commutation relations:
$$
[P^{\mu},x^\nu] \;\;  =\;\;  -i\eta^{\mu\nu}
$$
\noindent
Every space $H_p$
is generated by the infinite algebra
of the excitation operators:
\beqa
[\alpha_m^\mu,\alpha_n^\nu] & = & m\eta^{\mu \nu}\delta_{m,-n} \;\;\;,\cr
[\beta_m,\beta_n] & = & m\delta_{m,-n} ; \;\;\;,\;\;\;
m,n\in {Z\!\!\!Z}\;\;\;,
\label{ccr}
\enqa
out of the unique vacuum state $\Omega_p$
satisfying
\begin{eqnarray*}
\alpha_m^\mu\Omega_p &=&
\beta_m\Omega_p\;=\; 0\;\;\;,\;\;\;m > 0
\;\;\;,\\
\alpha_0^\mu\Omega_p &=&
{p^\mu\over \sqrt{\alpha}}\Omega_p
\;\;\;,\;\;\;
\beta_0\Omega_p \;=\;
q\Omega_p\;\;\;,
\end{eqnarray*}
where  the convention $\al_0^\mu :={1\over \sqrt{\alpha}} P^\mu $
is used. The eingenvalue $q$
of the operator $\beta_0$ is regarded as a free parameter of
the construction and is the same for all $\Omega_p$.
The scalar product in
$H_p$ is induced by (\ref{ccr}) and
the conjugation properties
$${(\alpha_m^\mu)}^\dagger = \alpha_{-m}^\mu~,~(\beta_m)^\dagger = \beta_{-m}
\;\;\;.$$

There is a realization of the Poincare algebra on $H$
with the generators of translations and Lorentz rotations
given by:
\beq
P^{\mu}\;\;\;,\;\;\;{\rm and}\;\;\;M^{\mu \nu}  =  P^\mu x^\nu -P^\nu x^\mu  + i\sum_{n=1}^{+\infty}
\frac {1}{n}(\alpha_{-n}^\mu \alpha_n^\nu - \alpha_{-n}^\nu \alpha_n^\mu )
\;\;\;,
\label{lo}
\enq
respectively.

One introduces the infinite set of quantum constraints
\beq
L_n  =\frac{1}{2}\sum_{m=-\infty}^{+\infty}  : \alpha_{-m}\cdot \alpha_{n+m} :
+ \frac{1}{2}\sum_{m=-\infty}^{+\infty} : \beta_{-m}\beta_{n+m} :
  +  2{\sqrt\beta}in\beta_n + 2\beta \delta_{n,0} \;\;\;,
\label{constra}
\enq
satisfying the Virasoro algebra:
$$
[L_n,L_m] = (n-m)L_{n+m} +{\textstyle {D+1+48\beta\over 12}} (n^3 - n)
\delta_{n,-m}\;\;\;.
$$
The space of physical states is defined as
$$
{\cH}=\{ \Psi \in H :\; (L_n-\delta_{n,0}a_0)\Psi = 0\;  ,
n \geq 0 \} \;\;\;.
$$
The $L_0$ operator is a combination of the momentum and
level operator $R$ and the corresponding constraint
yields the mass shell condition:
\beq
\left(     L_0-a_0\right) \Psi\;
= \; \left(
{\textstyle {1\over 2\alpha }}P^2 + R -\alpha(0)
\right)\Psi \;=\;0\;\;\; ,
\label{mshell}
\enq
where
 $
 \alpha(0) = a_0 -{\textstyle {1\over 2}}{q}^2
 - 2\beta
 $
is the physical intercept of leading Regge trajectory.

In the massive string model \cite{wy,my} one has an extra constraint
$\beta_0 \Psi =0$. In the present paper we relax this condition
keeping the arbitrary real eigenvalue $q$
of the operator $\beta_0$ as a free parameter.
In contrast to the parameters $\beta$ and $a_0$
which are restricted by the no-ghost theorem
$q$ may take arbitrary real value.

Since the Poincare generators (\ref{lo}) commute with the constraints
(\ref{constra}) they define  a representation
 of the Poincare algebra on ${\cH}$.
It follows from the mass shell condition (\ref{mshell}) that
the decomposition of ${\cH}$
into representations of a fixed mass coincides with the
level structure:
$$
{\cH} \;= \; \bigoplus_{N \geq 0} {\cH}^{(N)}\;\; ;\;\;\;
R\;{\cH}^{(N)} = N{\cH}^{(N)}\;\;\;.
$$
Each subspace ${\cH}^{(N)}$ is further decomposed into
a direct integral of finite dimensional spaces
${\cH}^{(N)}(p)$ with fixed
on-shell momentum:
$$
{\cH}^{(N)}\;=\; \int_{{\cal S}_N}d\mu^N (p)~
{\cH}^{(N)}(p)
\;\;\;,
$$
where ${\cal S}_N$ denotes the mass-shell at level $N$
determined by the condition (\ref{mshell})
$-p^2=m^2= 2\alpha (N+
{\textstyle {1\over 2}}{q}^2
+2\beta-a_0)$, and $d\mu^N(p)$ denotes
the Lorentz invariant measure on ${\cal S}_N$.

In order to analyse the physical content of the model one
needs a tractable parameterization of the space ${\cH}$
of physical states.
A global one can be obtained by
an appropriate modification
\cite{wy}, \cite{my} of the standard DDF operators
\cite{ddf}, \cite{brower}.
The DDF construction  starts with fixing
a light-cone frame $\{k,k',e^1,...,e^{D-2}\}$.
The transverse $D-2$ coordinates
$p^i= e^i\cdot p\,,\,i=1, ... ,D-2$,  and the light-cone coordinate
$p^+ = k\cdot p$
provide a global regular parameterization of all mass shells
corresponding to positive mass square and
a singular one in the case of tachyon.
The light-cone  coordinate $p^- = k'\cdot p$ on ${\cal S}_N$
becomes dependent and equals to $p^- =
{1\over 2p^+}({\overline p}^2 + 2\alpha(N - \alpha(0))$.
Since the points with $p^+= 0$ form  a zero-measure subset
of the tachyonic mass shell
this singularity of light-cone coordinates
has no effect in the quantum theory.

The construction of an appropriate set of the DDF operators
for a given light-cone basis
is briefly presented in the appendix A 1.
Let us only remark that the definitions
(\ref{atran}), (\ref{abrow}), (\ref{aliou})
differ from the standard ones by the
replacement
$$
k \longrightarrow {\sqrt{\alpha}\over k\cdot P}k\;\;\;,\;\;\;
k' \longrightarrow {k\cdot P\over \sqrt{\alpha}}k'\;\;\;.
$$
Due to this slight modification
the DDF operators are well defined on the whole
space ${\cH}$, while
in the conventional constructions
the domain is restricted by the condition
$ { k\cdot P \over \sqrt{ \alpha} }=k\cdot \alpha_0 \in {Z\!\!\!Z}$.

There are $D$ families of DDF operators creating the physical states:
$D-2$ families $\{ A^i_m \}_{m\in{Z\!\!\!Z}}$ generating transverse
excitations (\ref{atran}), one family $\{ C_m\}_{m\in{Z\!\!\!Z}}$ generating
Liouville excitations (\ref{aliou}), and one family
$\{ B_m \}_{m\in {Z\!\!\!Z}}$ of shifted Brower
operators (\ref{abrowd}) corresponding to the
longitudinal ones. These operators satisfy diagonalized
algebra
\beqa
[A_m^i,A_n^j] & = & m\delta^{ij}\delta_{m,-n}\;\;\;, \cr
[C_m,C_n] & = & m\delta_{m,-n}\;\;\;, \cr
[B_n,B_m] & = & (n-m)B_{n+m}
        +  \textstyle{ {25 - D - 48\beta\over 12}}(n^3 - n)\delta_{n,-m}
\;\;\;.
\label{com}
\enqa
\noindent
with all remaining commutators being zero. It is convenient to
 introduce additional family $\{ F_m \}_{m\in{Z\!\!\!Z}}$ of operators
(\ref{efop}) satisfying the commutation relations \cite{brgo,my}
\begin{eqnarray}
[A_m^i,F_n] &=& [C_m,F_n] \;=
\;[F_m,F_n]\;=\;0\;\;\;,\nonumber\\
{[B_m,F_n]} &=& -nF_{n+m}\;\;\;,
\nonumber
\end{eqnarray}
In contrast to the DDF operators $F_m$ do not commute with the
constraints (\ref{constra}).
All the DDF and $F$ operators have
definite level with respect to the original covariant level operator 
(\ref{level}):
$$
[\; R \;, \; {D}_m ] \; = \; -m {D}_m,
$$
where ${ D}$ denotes any of $A^i ,B ,C, F$. The virtue of introducing
$F$ operators is
that for every $N\ge1$ and every
$p\in {\cal S}_N$ with $p^+\neq 0$,
all ordered monomials  of the DDF and $F$ operators
of level $N$
\begin{equation}
 \vartheta^N(A^i,B, C,F)
\Omega_{p+{\sqrt{\alpha}N\over k\cdot p}k}
\label{base}
\end{equation}
form  a basis in the subspace ${H}^{(N)}(p)\subset H$
of all states from $H$ with the on-shell momentum $p$ \cite{brgo,my}.
It follows in particular that the states  generated only by the DDF
operators exhaust all physical states.

Let us note that the condition $p^+=0$
can be satisfied only by the
tachyonic on-shell momentum, but even in this case
momenta with $p^+=0$ form
a zero-measure subset of the mass shell.
Hence the modified DDF construction
provides a global parameterization of the space ${\cH}$.

In the DDF parametrization of the space of physical states
the inner product structure of ${\cH}$
is completely determined by the $B$-sector of
the algebra (\ref{comm}).
The general discussion of the "no-ghost" theorem
was given in \cite{my}. Here we are only interested in
the critical model in which the subspace of null states
$
{\cH}_0 = \{ \Psi \in {\cH} \;: (\; \Psi , \cdot \;) = 0 \}
$
is largest possible. As it can be easily inferred from
the construction of the shifted Brower longitudinal
operator $B_0$ (\ref{abrowd}) and the algebra (\ref{com})
this is the case of the critical values of parameters
$
\beta = {25-D\over 24}, a_0=1.
$
For these values all states containing $B$-excitations are null.

Since physical states which differ by a null state carry the
same physical information the space of "true" physical
states is given by the quotient
${\cH}_{\rm ph}= {\cH}/{\cH}_0$.
The representation
 (\ref{lo}) of the Poincare algebra
on ${\cH}$
induces a representation
on ${\cH}_{\rm ph}$. For each Poincare generator $M$
on ${\cal H}$ one defines the corresponding generator $M_{\rm ph}$
by its action on equivalence classes
$ M_{\rm ph}[\Psi] = [M\Psi]$,
$[\Psi ] \in {\cH}_{\rm ph}$.The physical content of the quantum model is completely
determined by the decomposition of
$ {\cH}_{\rm ph}$ into unitary irreducible
representations.

\section{Light-Cone Gauge}

As it was mentioned in the introduction one way to parameterize
the quotient ${\cH}_{\rm ph}={\cH}/{\cH}_0$
is to consider a subspace
${\cH}_{\rm gauge}\subset {\cH}$ containing only one
element from each equivalence class in ${\cH}_{\rm ph}$.
Such a subspace can be regarded as a gauge slice for the
quantum symmetry acting on states in ${\cH}$
by shifts in the null direction ${\cH}_0$.

In this section we shall consider the light-cone gauge
which is defined as  the subspace ${\cH}_{\rm \scriptscriptstyle LC} \subset {\cH}$
of all physical states generated by the
the transverse $A^i$
(\ref{atran}), and
the Liouville $C$
(\ref{aliou}) DDF operators.
Since in the critical massive string model all DDF states
containing shifted Brower modes $B$ (\ref{abrowd})
are null, the subspace ${\cH}_{\rm \scriptscriptstyle LC}$ is a good gauge
slice. It defines a section $\Sigma : {\cH}_{\rm ph}
\rightarrow {\cH}$ of the fibration
${\cH} \rightarrow {\cH}/{\cH}_0={\cH}_{\rm ph}$. $\Sigma$ is an isomorphism of
Fock spaces. The generators of the representation of the Poincare
algebra induced by $\Sigma$ on ${\cH}_{\rm lc}$
can be expressed directly
in terms of generators $M$ (\ref{lo}) on ${\cH}$
\beq
{M}_{\rm \scriptscriptstyle LC} \Psi =
\pi_{\rm \scriptscriptstyle LC} M \Psi\;\;\;,
\label{gelc}
\enq
where $\Psi \in {\cH}_{\rm \scriptscriptstyle LC}$ and
$\pi_{\rm \scriptscriptstyle LC}$
is the projection onto the light-cone gauge slice
${\cH}_{\rm \scriptscriptstyle LC}$ defined in the
base of diagonal DDF operators
by neglecting all terms containing the shifted
Brower $B$-excitations (\ref{abrowd}).

We shall show that the critical massive string
is isomorphic with the noncritical light-cone string introduced
in Section 1. Our strategy to prove  this equivalence is to
identify the space of states $H_{\rm lc}$ of the noncritical
light-cone string  as a subspace of the (pseudo) Hilbert space $H$
defining the massive string model and to construct
an isomorphism ${\cH}_{\rm \scriptscriptstyle LC} \rightarrow H_{\rm lc}$
as a projection along null direction.

Let as  introduce an auxiliary subspace $H_{\rm aux}\subset H$
defined by
\begin{eqnarray}
H_{\rm aux} &= & \bigoplus_{N \geq 0} H_{\rm aux}^{(N)}\;\; \;,
\nonumber\\
H_{\rm aux}^{(N)}&=& \int_{{\cal S}_N}d\mu^N (p)~
H_{\rm aux}^{(N)}(p)
\;\;\;,
\label{auxss}
\end{eqnarray}
where ${\cal S}_N$ denotes the mass-shell at level $N$
defined by the equation $m^2=2\alpha(N+{1\over 2}q^2 -{D-1\over 24})$,
and $H_{\rm aux}^{(N)}(p)$ is the subspace of all states
generated out of the vacuum $\Omega_p$ by all polynomials
of level $N$  in the creation operators $\alpha^+,
\alpha^i$, and $\beta$. The subspace $H_{\rm aux}$
has non-negative inner product with a large subspace of null
states. All states containing $\alpha^+$-excitations are null.
If the "Liouville" momenta $q$ are the same in both models
one can identify
the space of states $H_{\rm lc}$ of the
noncritical light-cone string (\ref{ssss})
as the subspace of $H_{\rm aux}$ containing
all states generated only by $\alpha^i$, and $\beta$
operators. Let us stress that the states from
$H_{\rm lc}$ are not physical states from the point
of view of the massive string model,
$H_{\rm lc}\cap{\cH}=\{0\}$.
The auxiliary space $H_{\rm aux}$ intersects the space
of physical states ${\cH}$ along the light-cone slice
$
{\cH}_{\rm \scriptscriptstyle LC}
= H_{\rm aux} \cap {\cH}
$.
Indeed for states
$\Psi \in {\cH}^{(N)}_{\rm \scriptscriptstyle LC}(p)$ one gets
\beqa
\Psi
&=& \vartheta^N (A^i,C)~
\Omega_{p+{\sqrt{\alpha}N\over k\cdot p}k} \label{lcstates}\\
&=&
\vartheta^N(\alpha^i, \beta){\Omega_p} +
\vartheta^N_{\rm rest}(\alpha^+,\alpha^i, \beta){\Omega_p}
\;\in\; H_{\rm aux}^{(N)}(p) \nonumber
\;\;\;,
\enqa
where each term of the polynomial
$\vartheta^N_{\rm rest}(\alpha^+,\alpha^i, \beta)$
contains at least one creation operator $\alpha^+$.

The virtue of introducing the subspace $H_{\rm aux}$
is that it contains both
$H_{\rm lc}$ and ${\cH}_{\rm \scriptscriptstyle LC}$.
Moreover one can obtain one subspace from the other
by deformation in the
null direction which considerably simplifies calculations
of the induced Poincare generators.

Let $\pi_{\rm lc}:H_{\rm aux}\rightarrow
H_{\rm lc}$ be the projection on the subspace $H_{\rm lc}$
defined by neglecting all terms containing $\alpha^+$-excitations.
It follows from (\ref{lcstates}) that
\beq
\sigma_{\rm lc}\;:\; H_{\rm lc} \ni
\vartheta^N({\alpha}^i, \beta)
\Omega_p
\longrightarrow
\vartheta^N (A^i,C)~
\Omega_{p+{\sqrt{\alpha}N\over k\cdot p}k}
 \in {\cH}_{\rm \scriptscriptstyle LC}\;\;\;.
\label{izomo}
\enq
is a section of $\pi_{\rm lc}$ (
$\pi_{\rm lc}\circ\sigma_{\rm lc} ={\rm id}$).
The composition
$\pi_{\rm lc}\circ\Sigma : {\cH}_{\rm ph} \rightarrow H_{\rm lc}$
is an isomorphism of Hilbert spaces and induces on $H_{\rm lc}$
a representation of the Poincare algebra
with the generators
$$
\pi_{\rm lc}\circ\pi_{\rm \scriptscriptstyle LC}
\circ M \circ \sigma_{\rm lc}
$$
where $M$ are generators (\ref{lo}) on ${\cal H}$.
In order to compare this representation
with the light-cone string representation
introduced in Section 1 one has to calculate the generators
in terms of the initial conditions
with respect to the $x^+$-evolution.

Note that by construction all states in  $H_{\rm aux}$ (\ref{auxss})
have on-mass-shell momenta.  In particular for
states from $H_{\rm lc}^{(N)}(p)\subset H_{\rm aux}^{(N)}(p)$
the $x^+$-dependence is given by
\begin{equation}
\vartheta^N({\alpha}^i, \beta) \Omega_p(p^+,\overline p,x^+)
=\vartheta^N({\alpha}^i, \beta)
{\rm e}^{{ix^+\over 2p^+}(\overline{p}^2 + 2\alpha(N-\alpha(0)))}
\Omega_p(p^+,\overline p, 0)\;\;\;.
\label{xpdep}
\end{equation}
For every operator $A$ on
$H$ we define the operator
$\overline A$ acting on the space of initial conditions
$\overline{\Psi}(p^+,\overline p)
=\Psi(p^+,\overline p, x^+)_{|x^+=0}$
of states from $H_{\rm lc}$ by
$$
\overline A \,\overline \Psi
= \overline{
\pi_{\rm lc}\circ\pi_{\rm \scriptscriptstyle LC}
\circ \pi_{\rm ph}
\circ A\circ\sigma_{\rm lc} \Psi
}\;\;\;,
$$
where $\pi_{\rm ph}:H\rightarrow {\cH}$ is the projection on the
space of physical states defined in the base (\ref{base})
by neglecting all terms containing the $F$ operators.
For the generators of translation one gets
$$
\overline{P}^i = P^i\;\;\;,\;\;\;
\overline{P}^+ = P^+\;\;\;,\;\;\;
\overline{P}^- =
{\alpha\over P^+}\left(L_0^{\rm \scriptscriptstyle T}
+ L_0^{\rm \scriptscriptstyle L}-1\right)\;\;\;,
$$
where $L_n^{\rm \scriptscriptstyle T}$ and $L_n^{\rm \scriptscriptstyle T}$
are defined by (\ref{ltra}) and (\ref{lob}), respectively.
Since since for all $\alpha^\mu_m$ operators
 $\overline{ : \alpha^\mu_m \alpha^\nu_n : } =
: \overline{\alpha}^\mu_m\,\overline{\alpha}^\nu_n :$,
the Lorentz generators can be written as
\beq
\overline{M}^{\mu \nu}  =
\overline{P}^\mu \overline{x}^\nu
-\overline{P}^\nu \overline{x}^\mu  + i\sum_{n=1}^{+\infty}
\frac {1}{n}(\overline{\alpha}_{-n}^\mu \overline{\alpha}_n^\nu -
\overline{\alpha}_{-n}^\nu \overline{\alpha}_n^\mu )
\;\;\;.
\label{lol}
\enq
Calculating $\overline{\alpha}_n^\mu$ and $\overline{x}^\mu$ one obtains
\begin{eqnarray*}
\overline{x}^i &=& x^i\;\;\;,\;\;\;
\overline{x}^+ = 0\;\;\;,\;\;\;
\overline{x}^- = x^-\;\;\;,\\
\overline{\alpha}^i_n &=& \alpha^i_n\;\;\;,\;\;\;
\overline{\alpha}^+_n = 0\;\;\;,\;\;\;
\overline{\alpha}^-_n =
{\sqrt{\alpha}\over P^+}\left(L_n^{\rm \scriptscriptstyle T}
+ L_n^{\rm \scriptscriptstyle L}-1\right)\;\;\;.
\end{eqnarray*}
Substituting in (\ref{lol}) one gets the formulae
(\ref{rot}), which completes the proof of equivalence.
Our considerations can be summarized in the following statement.
\bigskip

{\it The critical massive string model
with the "Liouville momenta" $q$ is isomorphic to
the noncritical light-cone string model with the
physical intercept $\alpha(0) = {D-1\over 24}-{1\over 2} q^2$.}
\bigskip

The light-cone slice ${\cH}_{\rm lc} \subset {\cH}$
can be also defined as the subspace of
all physical states $\Psi$ satisfying the quantum
light-cone gauge conditions
$k\cdot \al_n \Psi = \al_n^+\Psi =0\,,\, n>0,$
known from the critical Nambu-Goto string.
Let us stress that in the case of massive string one cannot
impose these gauge conditions in the classical theory, because
the Poisson bracket algebra of classical
constraints develops a central extension \cite{my}.
On the quantum level the critical massive model is the
only one where these conditions can be consistently
imposed to remove redundancy related to the null states.

\section{Nambu-Goto Gauge}

Let us consider the subspace
${\cH}_{\rm \scriptscriptstyle NG}\subset {\cH}$
consisting of all excited physical states generated by
the transverse
$ A^i$ (\ref{atran}) and
the longitudinal Brower
${\widetilde{B}}$ (\ref{abrow})
DDF operators. The states from ${\cH}_{\rm \scriptscriptstyle NG}  $
do not contain excitation in the Liouville sector of the model.
One can diagonalize the subalgebra generated by $A^i$ and $\widetilde{B}$
by introducing \cite{brower}
\beq
B^{\rm \scriptscriptstyle NG}_n
= \widetilde{B}_n - {\cal L}_n(A^i) + \delta_{n,0} \;\;\;,
\label{browng}
\enq
with
$$
{\cal L}_n(A^i)= {\textstyle{1\over 2}}\sum_{k=-\infty}^{+\infty}
:A^i_{-k}A^i_{n+k}:\;\;\;.
$$
The space ${\cH}_{\rm \scriptscriptstyle NG}$ is isomorphic
as a tensor product of Fock space (generated by $A^i$)
and Verma module (generated by $B^{\rm \scriptscriptstyle NG}$)
with the Hilbert space of the noncritical Nambu-Goto
string \cite{brower} with the intercept of leading Regge trajectory
$\alpha(0)={D-1\over 24}-{1\over 2}q^2$.

Using the momenta and level decomposition of the space
of excited physical states and the algebra (\ref{addf})
one can prove by the standard counting argument
that ${\cH}_{\rm \scriptscriptstyle NG}$ is a good gauge slice.
Another more geometrical proof consists in
constructing ${\cH}_{\rm \scriptscriptstyle NG}$
by a transformation of the light-cone slice
along null direction.
The space ${\cH}_{\rm \scriptscriptstyle LC}$ has the structure
of the Fock space
generated by the transverse $A^i$
and the Liouville $C$ excitations.
There is an equivalent description of this space
 as a tensor product of the Fock space generated by
the  transverse excitations and the Verma module
generated by the Virasoro algebra constructed
from the Liouville modes  \cite{nove}:
$$
{\cal L}_n(C) = + \frac{1}{2} \sum_{m=-\infty}^{+\infty} : C_{-m}C_{n+m} :
 + 2i{\sqrt\beta}nC_n + 2\beta\delta_{n,0}\;\;\;.
$$
Any state from ${\cH}_{\rm \scriptscriptstyle LC}^{(N)}(p)$ can be written as an element
of the Verma module created out of the highest weight vacuum vector :
$$
{\cH}_{\rm \scriptscriptstyle LC}^{(N)}(p) \ni \Psi =
\vartheta^N({\cal L}(C),A)\Omega_{p+{\sqrt{\alpha}N\over k\cdot p}k}
\;\;\;.
$$
Let us notice that the Brower
${B}^{\rm\scriptscriptstyle NG}_{n}$ (\ref{browng})
DDF operators     differ from
the Virasoro generators ${\cal L}_n(C)$
 by the shifted  Brower modes $B_n$ (\ref{abrowd})
$$
  {B}^{\rm\scriptscriptstyle NG}_{n} - {\cal L}_n(C)
 = B_n \;\;\;,\;\;\;n\in Z\!\!\!Z\;\;\;.
$$
Since $[ B_n,{\cal L}_n(C)] = 0$ and the central charge of the shifted
Brower modes is zero (\ref{com}),
the map defined level by level by
\beq
{\cH}^{(N)}_{\rm \scriptscriptstyle LC}(p) \ni
\vartheta^N({\cal L}(C),A)
\Omega_{p+{\sqrt{\alpha}N\over k\cdot p}k}
\mapsto \vartheta^N(B^{\rm\scriptscriptstyle NG},A)
\Omega_{p+{\sqrt{\alpha}N\over k\cdot p}k}
\in {\cH}^{(N)}_{\rm \scriptscriptstyle NG}(p) \;\;\;,
\label{shift}
\enq
is an isomorphism of Hilbert spaces preserving the structure
of tensor product of Fock space and Verma module.
For each state
in ${\cH}_{\rm \scriptscriptstyle LC}$ the map (\ref{shift}) is a shift
in null direction. The image ${\cH}_{\rm \scriptscriptstyle NG}$
is therefore a good gauge slice.

In contrast to the light-cone gauge
${\cH}_{\rm \scriptscriptstyle LC}$ the subspace ${\cH}_{\rm \scriptscriptstyle NG}$
is stable with respect to the Poincare transformations
in ${\cH}$.  The induced representation of the Poincare
algebra is simply given by the restriction of the
generators $M$ (\ref{lo}) on ${\cH}$ to the subspace
${\cH}_{\rm \scriptscriptstyle NG}$ and is identical with the
representation one obtains in the covariant quantization of
noncritical Nambu-Goto string with
the physical intercept  $\alpha(0)={D-1\over 24} -{1\over 2}q^2$.
One gets therefore the following result:
\bigskip

{\it The critical massive string model
with the "Liouville momentum" $ q$ is equivalent
to the non-critical Nambu-Goto string model with
the physical intercept $\alpha(0) = {D-1\over 24}-{1\over 2}q^2$.}
\bigskip
\medskip

\section*{Acknowledgements}

The authors would like to thank Andrzej Ostrowski for
many discussions, and for help with the numerical calculations.
We would also like to thank
Jurek Cis{\l}o for enlightening discussion
on generating functions.
This work is supported in part by
the Polish Committee of Scientific Research
(Grant Nr PB 1337/PO3/97/12).

\section*{ Appendix }
\renewcommand{\theequation}{A.\arabic{equation}}
\setcounter{equation}{0}
\noindent
In order to define the DDF operators one introduces the fields:
\beqa
X^{\mu}(\theta) & = & q_0^{\mu} + \alpha_0^{\mu}\theta + 
\sum_{m\ne0}\frac{i}{m}\alpha_m^{\mu}{\rm e}^{-im\theta}
\;\;\;, \cr
\Phi(\theta) & = & \sum_{m\ne0}\frac{i}{m}\beta_m{\rm e}^{-im\theta}
\;\;\;,\cr
P^{\mu}(\theta) & = & X^{\mu'}(\theta) = \sum_{m=-\infty}^{+\infty} \alpha_m^{\mu}
{\rm e}^{-im\theta}
\;\;\;,\cr
\Pi(\theta) & = & \Phi'(\theta) = \sum_{m=-\infty}^{+\infty} \beta_m{\rm e}^{-im\theta}
\nonumber\;\;\;,
\enqa
where $q_0^\mu = {\sqrt\alpha}x^\mu , \al_0^\mu =\frac{1}{\sqrt{\alpha}}P^\mu$ ,
with the following commutation rules:
\begin{eqnarray}
{[X^\mu(\theta),X^\nu(\theta')]} & = &
-2\pi i \eta^{\mu\nu}\epsilon(\theta-\theta') 
\;\;\;,\nonumber \\
{[\Phi(\theta),\Phi(\theta')]} & = &
-2\pi i (\epsilon(\theta-\theta') + \epsilon
(\theta'-\theta))
\;\;\;,\nonumber \\
{[P_\mu(\theta),X^\nu(\theta')]} & =&
-2\pi i \delta_\mu^\nu \delta (\theta-\theta')
\;\;\;,\nonumber\\
{[\Pi(\theta),\Phi(\theta')]} & = &
-2\pi i (\delta(\theta-\theta') - 1)
\;\;\;,\nonumber \\
{[P_\mu(\theta),P_\nu(\theta')]} & = &
-2\pi i \eta_{\mu\nu}\delta'(\theta-\theta')
\;\;\;, \nonumber \\
{[\Pi(\theta),\Pi(\theta')]} & = & -2\pi i \delta'(\theta-\theta')
\;\;\;.\nonumber
\end{eqnarray}
The definitions of the transverse
\beq
A_m^i(k) = \frac{1}{2\pi}\int\limits_0^{2\pi} d\theta
: e_i\cdot P(\theta){\rm e}^{imkX(\theta)\over k\cdot \alpha_0}:
\;\;\;  ,\;\;\; i \; = \; 1,...,D-2\;\;\;.
\label{atran}
\enq
and of the longitudinal
\beq
{\widetilde B}_m = \frac{1}{2\pi}\int\limits_0^{2\pi}d\theta
:\left( k\cdot \alpha_0 \; k'\cdot P(\theta)
- \frac{im}{2}{\rm log}'(\frac{k\cdot P(\theta)}{k\cdot \al_0})\right) 
{\rm e}^{imkX(\theta)\over k\cdot \alpha_0}:\;\;\;.
\label{abrow}
\enq
DDF operators are only slight modifications of the standard constructions
of \cite{ddf}, and \cite{brower}.
The "Liouville" DDF operators corresponding to
the $\beta$ excitations are defined by \cite{wy}:
\beq
C_m = \frac{1}{2\pi}\int\limits_0^{2\pi}d\theta :\left(\Pi(\theta) - 
2{\sqrt\beta}{\rm log}'(\frac{k\cdot P(\theta)}{k\cdot \al_0})\right) 
{\rm e}^{imkX(\theta)\over k\cdot \al_0 }: \;\;\;.
\label{aliou}
\enq
The additional family of $F$ operators generating non-physical
states is defined by \cite{brgo}
\begin{equation}
F_m(k) = {1\over 2\pi} \int\limits_0^{2\pi} d\theta\,
:{\rm e}^{imk\cdot X(\theta)}: \;\;\;.
\label{efop}
\end{equation}
In contrast to the DDF operators $F_m(k)$ do not commute with the
constraints
\begin{eqnarray}
\nonumber
[L_m,F_n(k)]& =& -mF_n^m(k)
\;\;\;,\\
F_m^n(k) &=& {1\over 2\pi}
\int\limits_0^{2\pi} d\theta\,{\rm e}^{in\theta}
:{\rm e}^{imk\cdot X(\theta)}: \;\;\;.\nonumber
\end{eqnarray}
The algebra of the DDF operators reads:
\begin{eqnarray}
[A_m^i(k),A_n^j(k)] & = & m \delta^{ij} \delta_{m,-n}\;\;\;,
\nonumber\\
{[C_m(k),C_n(k)]} &=& m\delta_{m,-n}\;\;\;,
\nonumber\\
{[\widetilde{B}_m(k),\widetilde{B}_n(k)]} & = & 
(n-m) \widetilde{B}_{n+m}(k) + 
2n^3\delta_{m,-n}
\;\;\;,
\label{addf}\\
{[\widetilde{B}_n(k), A_m^i(k) ]} &=& -m A^i_{m+n}(k) \;\;\;,
\nonumber\\
{[\widetilde{B}_n(k), C_m(k)]} & =& 
-mC_{n+m}(k)+2in^2\sqrt{\beta} \delta_{n,-m}\;\;\;.
\nonumber
\end{eqnarray}
The algebra above can be diagonalised by
introducing the shifted Brower operators  \cite{brower}, \cite{my}:
\beq
{\widetilde B}_n\longrightarrow B_n = {\widetilde B}_n
- {\cal L}_n + \delta_{n,0}
\label{abrowd}
\enq
where
\beq
{\cal L}_n = \frac{1}{2} \sum_{m=-\infty}^{+\infty}
:A_{-m}^i\cdot A_{n+m}^i : + \frac{1}{2}
\sum_{m=-\infty}^{+\infty} : C_{-m}
 C_{n+m}:
 + 2i{\sqrt\beta}C_n + 2\beta\delta_{n,0}.
 \label{al}
\enq
For the new basis $\left\{A^i,B,C\right\}$ of
the DDF operators one gets the "diagonal" commutation relations
(\ref{com}).

\end{document}